# New Journal of Physics

The open access journal at the forefront of physics

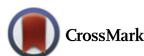

Published in partnership with: Deutsche Physikalische Gesellschaft and the Institute of Physics

# Local model of a qubit in the interferometric setup

### Pawel Blasiak

Institute of Nuclear Physics, Polish Academy of Sciences, ul. Radzikowskiego 152, 31-342 Kraków, Poland

**E-mail:** pawel.blasiak@ifj.edu.pl

## Abstract

We consider a typical realization of a qubit as a single particle in two-path interferometric circuits built from phase shifters, beam splitters and detectors. This framework is often taken as a standard example illustrating various paradoxes and quantum effects, including non-locality. In this paper we show that it is possible to simulate the behaviour of such circuits in a classical manner using stochastic gates and two kinds of particles, *real* ones and *ghosts*, which interact only locally. The model has built-in limited information gain and state disturbance in measurements which are blind to *ghosts*. We demonstrate that predictions of the model are operationally indistinguishable from the quantum case of a qubit, and allegedly 'non-local' effects arise only on the epistemic level of description by the agent whose knowledge is incomplete due to the restricted means of investigating the system.

## 1. Introduction

Quantum mechanics presents a challenge to many classical concepts that we hold about the world. In particular, it defies the very essence of particle ontology which says that a particle is localized only in one place at a time and interacts only with objects in its immediate vicinity. It was a profound insight of Bell [1–3] to point out that quantum mechanics admits correlations between particles which contradict the assumption of local realism. As a consequence, to recover quantum predictions in a realistic hidden-variable model one has to resort to spooky action at a distance and thus violate the paradigm of locality. In a similar manner it is often argued that in the case of a *single particle* a kind of non-local influence is also required to account for the effects associated with the collapse of the wave function. For an illustration of this type of reasoning it is enough to consider simple interferometric setups, e.g. see the analysis of single-particle interference in the Mach–Zehnder interferometer [4, 5] or the proposal of interaction-free measurements [6–8]. These sorts of arguments exploit the apparent difficulty in answering the question: *how does the particle, being localized in a given path, know what happens in the other path of the interferometer?* Faced with a puzzle, conventional wisdom attributes this kind of behaviour to non-local effects—either of the particle itself or the wave function. However, it is unclear if this is enough to establish similar conclusions as in the Bell-type reasoning. In particular, does it imply the impossibility of local hidden variable models simulating quantum behaviour in the considered interferometric setups? In this paper, we answer this question in the negative.

Clearly, in the Bell scenario one is concerned with correlations between a pair of quantum particles, whereas in the single particle case we are concerned with a single quantum particle interacting with classical apparatus. Hence the question of the hidden variable account is brought up in a different conceptual context. This makes it interesting to ask if it is possible to simulate the single-particle behaviour of interferometric circuits by replacing quantum gates with stochastic counterparts without violating the locality principle. Note that an argument of the Bell-type does not apply in this situation, and hence it should not be very surprising if a different conclusion is reached.

In this paper, we take a closer look at a single-particle framework for two-path interferometric setups built from phase shifters, beam splitters and detectors. It has a simple description which boils down to a qubit and, as such, is often taken as the prototypical example illustrating various paradoxes and quantum effects, see e.g. [4–14]. We show that it is possible to simulate the behaviour of such circuits in a classical manner using stochastic gates and particles which interact only locally. The crucial ingredient of the model is the existence of two kinds of particles, *real* ones and *ghosts*, with the latter being invisible to detectors. This allows one to construct a stochastic analogue of quantum circuits with built-in limited information gain and state disturbance. We show that the operational description of the system—by an agent investing the system according to the rules of the model—parallels the quantum case, i.e. boils down to a qubit. At the same time, on the ontological level all gates and particles considered in the model conform to the paradigm of locality.

The paper is organized as follows. We start with an informal introduction of the main concepts in section 2. Section 3 gives a concise account of the quantum-interferometric building blocks and their description in terms of a qubit. In section 4 we explain the ontology of the model, discuss the criterion of locality and define stochastic analogues of the interferometric gates. The model is carefully analysed in section 5 where we start from the ontic description of epistemic constraints imposed on the agent investigating the system, and then abstract away all unnecessary details to give a purely operational account of the model as seen by the agent unaware of the underlying ontology. The latter is shown to be equivalent to the description of a qubit. We conclude with a brief discussion in section 6.

## 2. Heuristics of the model

Our main goal in this paper is the construction of a local stochastic model which simulates the behaviour of a quantum particle in every conceivable two-path interferometric circuit. Before going into the details of the model it is instructive to give a glimpse of how it works on a few simple variants of the Mach–Zehnder interferometer. In this section we give an informal discussion of a toy version of the model which will be generalized and carefully analysed throughout the paper.

Let us consider a single quantum particle entering a circuit which consists of two paths labelled $i = 0$ and 1, two 50:50 beam splitters $B$ with a $\pi$-phase shifter $P_0(\pi)$ or a detector $D_0$ in-between and two detectors $D_0$ and $D_1$ at the end measuring the statistics. See the circuits illustrated in figure 2. A quantum description of the system which starts in state $|0\rangle$ (i.e. a particle in the upper path $i = 0$) boils down to the following sequence of transformations (with the convention that a particle upon reflection on the beam splitter gains phase i, and otherwise goes through unaffected). Without a phase shifter we have

$$|0\rangle \xrightarrow{B} \tfrac{i}{\sqrt{2}}(|0\rangle - i|1\rangle) \xrightarrow{B} i|1\rangle, \tag{1}$$

i.e. the particle is always detected in the lower path $i = 1$. With a $\pi$-phase shifter we obtain

$$|0\rangle \xrightarrow{B} \tfrac{i}{\sqrt{2}}(|0\rangle - i|1\rangle)$$
$$\xrightarrow{P_0(\pi)} \tfrac{-i}{\sqrt{2}}(|0\rangle + i|1\rangle) \xrightarrow{B} |0\rangle, \tag{2}$$

and the particle always ends in the upper path $i = 0$. For a detector placed in the upper path we have the following description

$$|0\rangle \xrightarrow{B} \tfrac{i}{\sqrt{2}}(|0\rangle - i|1\rangle) \xrightarrow{D_0} \begin{cases} |0\rangle & \text{`Click'} \\ |1\rangle & \text{`No Click'} \end{cases}$$
$$\xrightarrow{B} \begin{cases} \tfrac{i}{\sqrt{2}}(|0\rangle - i|1\rangle) \\ \tfrac{1}{\sqrt{2}}(|0\rangle + i|1\rangle), \end{cases} \tag{3}$$

which means that the particle is detected with equal probability in either path. The question is whether it is possible to simulate this behaviour in a local manner using stochastic gates.

To build a toy model of the phenomenon let us consider two kinds of particles which differ by the response of detectors. Namely, particles of the first kind always trigger the detector ('Click') while particles of the second kind leave the detector without response ('No Click'). We will call them metaphorically *real* and *ghost* particles respectively. Additionally, we will assume that the particles have inner degrees of freedom: for the *real* particle it is a unit vector pointing in the direction $\vec{n} = \pm \hat{y}$ or $\pm \hat{z}$ in 3D space, and for the *ghost* particle it is a phase $\varphi = 0$ or $\pi$; see figure 1. If we restrict our attention to a single *real* particle in the system which is possibly accompanied by a *ghost* in the other path, then we can encode this situation by a triple

$$(i, \vec{n}, \varphi) \quad \text{or} \quad (i, \vec{n}, \varnothing),$$

where $i = 0, 1$ labels the path with the *real* particle, $\vec{n}$ is its inner state, and in the first case $\varphi$ is the phase of the *ghost* while in the second case $\varnothing$ indicates that the other path is *empty*. To build a model we need to specify how these states transform under the action of the stochastic counterparts of the interferometric gates.

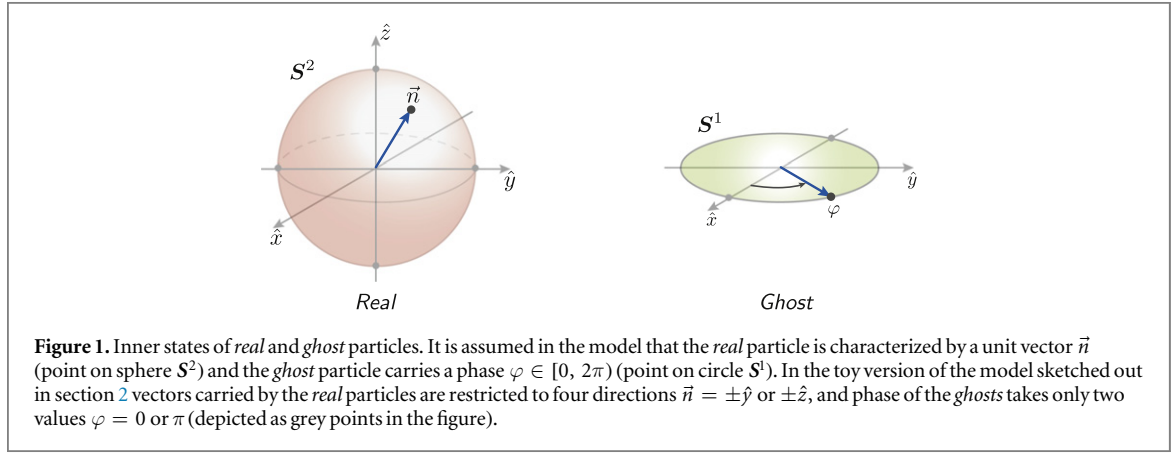

**Figure 1.** Inner states of *real* and *ghost* particles. It is assumed in the model that the *real* particle is characterized by a unit vector $\vec{n}$ (point on sphere $S^2$) and the *ghost* particle carries a phase $\varphi \in [0, 2\pi)$ (point on circle $S^1$). In the toy version of the model sketched out in section 2 vectors carried by the *real* particles are restricted to four directions $\vec{n} = \pm\hat{y}$ or $\pm\hat{z}$, and phase of the *ghosts* takes only two values $\varphi = 0$ or $\pi$ (depicted as grey points in the figure).

Let us first describe a *beam splitter* $\mathbb{B}$. Imagine that we have a *real* and *ghost* particle in each path, i.e. state $(i, \vec{n}, \varphi)$. Then the beam splitter either leaves $(i \to i)$ or swaps $(i \to \bar{\imath} \equiv 1 - i)$ particles in their respective paths and changes their inner state according to the following prescription

$$(i, \vec{n}, \varphi) \xrightarrow{\mathbb{B}} \begin{cases} (i, \vec{n}', 0) & \text{with probability } \cos^2\left(\frac{\theta'}{2}\right), \\ (\bar{\imath}, -\vec{n}', 0) & \text{with probability } \sin^2\left(\frac{\theta'}{2}\right), \end{cases} \qquad (4)$$

where $\vec{n}' = (\theta', \phi') = R_x\left(\frac{\pi}{2}\right) R_z(-\varphi)\vec{n}$. Note that to implement the gate one requires information from both paths together and the respective probabilities can take values $0, \frac{1}{2}$ and $1$ depending on the inner states of both particles $\vec{n}$ and $\varphi$. This means that the gate is *non-local* and *stochastic*. More explicitly, we get with equal probability

$$(i, \pm\hat{z}, \varphi) \begin{array}{c} \xrightarrow{\frac{1}{2}} (i, \mp\hat{y}, 0) \\ \xrightarrow{\frac{1}{2}} (\bar{\imath}, \pm\hat{y}, 0) \end{array}$$

for $\varphi = 1, \pi$, and with certainty

$$(i, +\hat{y}, 0) \longrightarrow (i, +\hat{z}, 0), \qquad (i, +\hat{y}, \pi) \longrightarrow (\bar{\imath}, +\hat{z}, 0),$$
$$(i, -\hat{y}, 0) \longrightarrow (\bar{\imath}, +\hat{z}, 0), \qquad (i, -\hat{y}, \pi) \longrightarrow (i, +\hat{z}, 0).$$

To complete the description we should also specify action of the beam splitter $\mathbb{B}$ when one of the paths is *empty*, i.e. state $(i, \vec{n}, \varnothing)$. Then we define

$$(i, \vec{n}, \varnothing) \xrightarrow{\mathbb{B}} \begin{cases} (i, -\hat{y}, 0) & \text{with probability } \frac{1}{2}, \\ (\bar{\imath}, +\hat{y}, 0) & \text{with probability } \frac{1}{2}, \end{cases} \qquad (5)$$

which means that the beam splitter creates a *ghost* particle in the other path.

For a $\pi$-*phase shifter* $\mathbb{P}_0(\pi)$ placed in the upper path we make the following definition

$$(0, \vec{n}, \varphi) \xrightarrow{P_0(\pi)} (0, R_z(-\pi)\vec{n}, \varphi),$$
$$(1, \vec{n}, \varphi) \xrightarrow{P_0(\pi)} (1, \vec{n}, \varphi + \pi), \qquad (6)$$

and

$$(0, \vec{n}, \varnothing) \xrightarrow{P_0(\pi)} (0, R_z(-\pi)\vec{n}, \varnothing),$$
$$(1, \vec{n}, \varnothing) \xrightarrow{P_0(\pi)} (1, \vec{n}, \varnothing). \qquad (7)$$

In other words, the gate takes the particle in the upper path and rotates around the $\hat{z}$-axis by angle $-\pi$ if it is the *real* particle, or by angle $+\pi$ if it is the *ghost* (for the *empty* path it does nothing). Note that this is a *deterministic* gate which acts *locally* in the upper path (i.e. it does not affect nor require any information from the other path).

Finally, we assume that *detectors* in the model are sensitive ('CLICK') only to *real* particles, and remain silent about the *ghosts* ('NO CLICK'). Additionally, we postulate that after detection the inner state of the *real* particle is set to $\vec{n} \to +\hat{z}$ while the *ghosts* which hit the detector are removed from the path $\varphi \to \varnothing$. In other words, for a detector $\mathbb{D}_i$ placed in the $i$th path we define

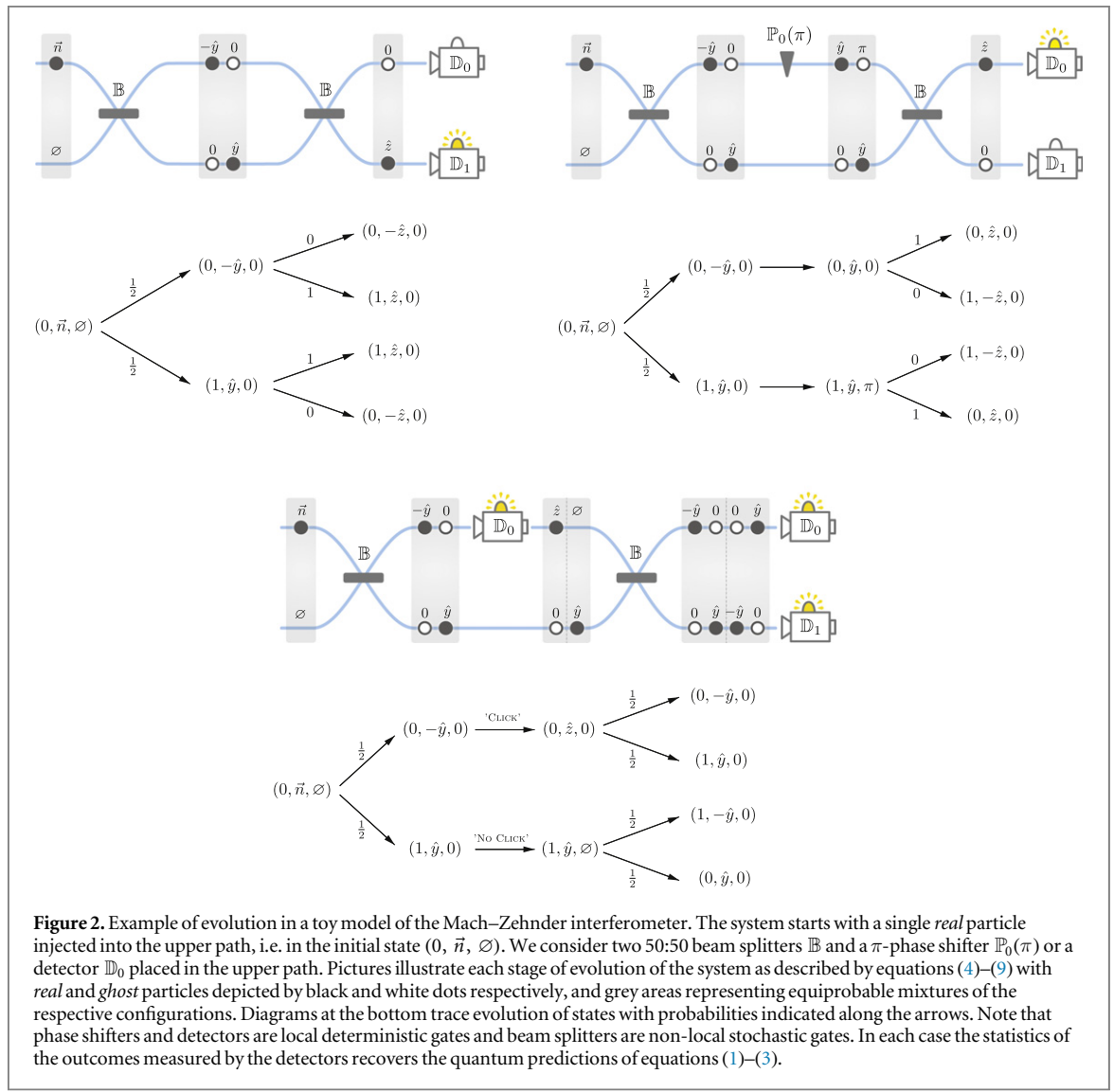

**Figure 2.** Example of evolution in a toy model of the Mach–Zehnder interferometer. The system starts with a single *real* particle injected into the upper path, i.e. in the initial state $(0, \vec{n}, \varnothing)$. We consider two 50:50 beam splitters $\mathbb{B}$ and a $\pi$-phase shifter $\mathbb{P}_0(\pi)$ or a detector $\mathbb{D}_0$ placed in the upper path. Pictures illustrate each stage of evolution of the system as described by equations (4)–(9) with *real* and *ghost* particles depicted by black and white dots respectively, and grey areas representing equiprobable mixtures of the respective configurations. Diagrams at the bottom trace evolution of states with probabilities indicated along the arrows. Note that phase shifters and detectors are local deterministic gates and beam splitters are non-local stochastic gates. In each case the statistics of the outcomes measured by the detectors recovers the quantum predictions of equations (1)–(3).

$$(i, \vec{n}, \varphi) \xrightarrow[\text{'CLICK'}]{\mathbb{D}_i} (i, \hat{z}, \varphi), \qquad (\bar{i}, \vec{n}, \varphi) \xrightarrow[\text{'No CLICK'}]{\mathbb{D}_i} (\bar{i}, \vec{n}, \varnothing), \tag{8}$$

and

$$(i, \vec{n}, \varnothing) \xrightarrow[\text{'CLICK'}]{\mathbb{D}_i} (i, \hat{z}, \varnothing), \qquad (\bar{i}, \vec{n}, \varnothing) \xrightarrow[\text{'No CLICK'}]{\mathbb{D}_i} (\bar{i}, \vec{n}, \varnothing). \tag{9}$$

Clearly, detectors defined in this way are *local deterministic* gates.

Now, we can check the behaviour of the Mach–Zehnder interferometer built from the stochastic gates defined above. Figure 2 illustrates the evolution of the system with a single *real* particle injected into the upper path, i.e. starting from the initial state $(0, \vec{n}, \varnothing)$. In all three cases the toy model gets the quantum statistics right, see equations (1)–(3). This also turns out to be true for any longer sequence of gates one could build and another initial state $(1, \vec{n}, \varnothing)$. A closer look at the evolution of the system reveals a pattern in the possible states that can be generated along the way which can be grouped into the following four disjoint classes

$$[+\hat{z}]: \left\{ (0, \hat{z}, \varphi), (0, \vec{n}, \varnothing) \right\},$$

$$[-\hat{z}]: \left\{ (1, \hat{z}, \varphi), (1, \vec{n}, \varnothing) \right\},$$

$$[+\hat{y}]: \left\{ (0, \hat{y}, 0) \vee (1, -\hat{y}, \pi), (0, -\hat{y}, \pi) \vee (1, \hat{y}, 0) \right\},$$

$$[-\hat{y}]: \left\{ (0, -\hat{y}, 0) \vee (1, \hat{y}, \pi), (0, \hat{y}, \pi) \vee (1, -\hat{y}, 0) \right\},$$

where the connective '$\vee$' means an equiprobable mixture of the respective states. There are six states in each class $[\pm\hat{z}]$ and two probabilistic mixtures in each class $[\pm\hat{x}]$ and $[\pm\hat{y}]$. We note that these distributions have non-overlapping supports.

Without going into a detailed analysis we make the following observations about the classes of states defined above. Firstly, these are all possible states that can be generated from the initial states $(i, \vec{n}, \varnothing) \in [\pm\hat{z}]$. Secondly, statistics of measurement outcomes is the same for all states in a given class (i.e. for states in $[\pm\hat{z}]$ detectors in the respective paths 'CLICK'/'NO CLICK' with certainty, while for states in $[\pm\hat{y}]$ detectors 'CLICK' with equal probability and are anti-correlated). Thirdly, these classes of states transform congruently under the action of the stochastic gates defined above. This means that each gate transforms classes as the whole $[\vec{N}] \Rightarrow [\vec{N}']$ which defines a mapping of vectors labelling the classes $\vec{N} \rightarrow \vec{N}'$.[1] Furthermore, one can show that beam splitters $\mathbb{B}$ correspond to rotations about the $\hat{x}$-axis $[\vec{N}] \Rightarrow [R_x(\pi/2)\vec{N}]$, phase shifters $\mathbb{P}_0(\pi)$ generate rotations about the $\hat{z}$-axis $[\vec{N}] \Rightarrow [R_z(-\pi)\vec{N}]$, and detectors induce projections $[\vec{N}] \Rightarrow [\pm\hat{z}]$ with respective probabilities $\cos^2(\theta/2)$ and $\sin^2(\theta/2)$ determined by the polar angle of the labelling vector $\vec{N} = (\theta, \phi)$ where $\theta = 0, \frac{\pi}{2}, \pi$ and $\phi = 0, \pi$. This behaviour is nicely captured in the following diagrams

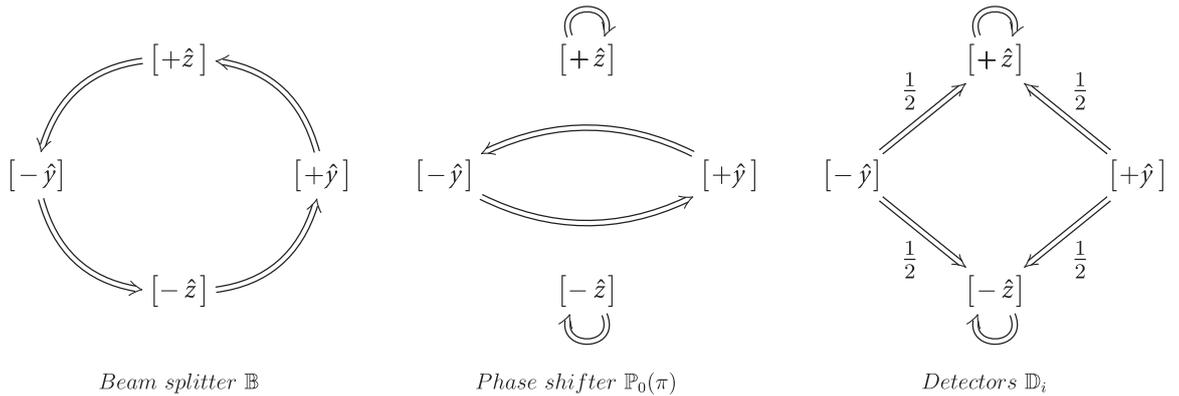

*Beam splitter* $\mathbb{B}$          *Phase shifter* $\mathbb{P}_0(\pi)$          *Detectors* $\mathbb{D}_i$

Note the resemblance to the geometric description of quantum interferometric gates in the Bloch representation of a qubit.

Now, consider an agent who is not interested or acquainted with the details of the toy model as described above. Suppose that the agent is only concerned with predicting the statistics of measurements in circuits built from any sequence of gates initiated by injecting *real* particles into one of the paths. In this situation, a purely operational account of the system is enough and all relevant information is encoded in the vector labelling the class (a detailed description of the states is irrelevant since only the statistics of 'CLICKS' matters which is the same for each class and the latter transform in a congruent manner). This allows one to make the following identification of classes with quantum states of a qubit

$$[+\hat{z}] \quad \longleftrightarrow \quad |+\hat{z}\rangle \equiv |0\rangle,$$
$$[-\hat{z}] \quad \longleftrightarrow \quad |-\hat{z}\rangle \equiv |1\rangle,$$
$$[+\hat{y}] \quad \longleftrightarrow \quad |+\hat{y}\rangle \equiv \tfrac{1}{\sqrt{2}}(|0\rangle + \mathrm{i}|1\rangle),$$
$$[-\hat{y}] \quad \longleftrightarrow \quad |-\hat{y}\rangle \equiv \tfrac{1}{\sqrt{2}}(|0\rangle - \mathrm{i}|1\rangle).$$

These quantum states share the same pattern of transformation rules for the corresponding gates (see the Bloch representation of a qubit). In conclusion, for all practical purposes the behaviour of circuits built from the stochastic gates in the toy model is indistinguishable from the corresponding quantum interferometric circuits whose description is restricted to the set of four states $|\pm\hat{y}\rangle$ and $|\pm\hat{z}\rangle$.

This section was meant as a heuristic introduction to the full model discussed in the following sections which recovers a complete description of a qubit. We have sketched out the key features of the model and illustrated the methodology for building an operational account of the system in which a full ontological description is irrelevant or inaccessible to the agent who is only interested in the statistical predictions of the model. It was indicated that in this restricted framework an operational account is equivalent to the description of a certain subset of states of a qubit. These ideas will be made precise in the following sections where the toy model is generalized to the full set of interferometric gates and a complete ontological analysis is carried out to show the operational indistinguishability of the model from the quantum description of a qubit.

---

[1] Note that we use upper case $\vec{N}$ to label classes $[\pm\hat{y}]$ and $[\pm\hat{z}]$ which should be distinguished from the lower case letter $\vec{n}$ associated with the inner state of the *real* particle.

## 3. Interferometric setup for a qubit

Let us set the stage for the discussion in the following sections by recalling a few basic facts about the Bloch representation of a qubit and its realization in the interferometric circuits.

A qubit is the simplest quantum mechanical system which is described in the Hilbert space $\mathcal{H} = \mathbb{C}^2$ [15]. It is convenient to represent *pure states* of a qubit

$$|\psi\rangle = \cos\frac{\theta}{2}\,|0\rangle + \mathrm{e}^{\mathrm{i}\phi}\sin\frac{\theta}{2}\,|1\rangle = \begin{pmatrix} \cos\frac{\theta}{2} \\ \mathrm{e}^{\mathrm{i}\phi}\sin\frac{\theta}{2} \end{pmatrix} \tag{10}$$

by points $\boldsymbol{n} = (\theta,\,\phi)$ on the so called *Bloch sphere*, where $\theta$ and $\phi$ are standard polar and azimuthal angles in 3D. *Mixed states* have a similar representation which extends to the *Bloch ball* via the parametrization $\rho = \frac{1}{2}(\mathbb{1} + \boldsymbol{n} \cdot \boldsymbol{\sigma})$, where $|\boldsymbol{n}| \leqslant 1$ (with pure states $|\psi\rangle\langle\psi|$ lying on the surface, i.e. $|\boldsymbol{n}| = 1$). In this representation unitary transformations $\rho \rightarrow U\rho U^{\dagger}$ (or $|\psi\rangle \rightarrow U\,|\psi\rangle$ for pure states) correspond to rotations $\boldsymbol{n} \rightarrow R_{\hat{r}}(\vartheta)\boldsymbol{n}$, where the axis $\hat{r}$ and the angle $\vartheta$ is determined from the parametrization $U = \mathrm{e}^{\mathrm{i}\alpha}\mathrm{e}^{-\mathrm{i}\vartheta\,\hat{r}\cdot\boldsymbol{\sigma}/2}$ (with $\alpha$ being an irrelevant overall factor). According to the Born rule measurement in the computational basis $|0\rangle \equiv \binom{1}{0}$ and $|1\rangle \equiv \binom{0}{1}$ (or equivalently $\pm\hat{z}$ in the Bloch repr.) on a system in state $\rho$ (or equivalently $\boldsymbol{n}$) yields outcome $i = 0, 1$ with probability

$$P_{\rho}(i) = \mathsf{Tr}\left(\rho\,|i\rangle\langle i|\right) = \frac{1}{2}(1 + (-)^{i}\,\hat{z} \cdot \boldsymbol{n}), \tag{11}$$

and leaves the system in the corresponding pure state

$$\rho \longrightarrow \begin{cases} |0\rangle, & \text{for } i = 0, \\ |1\rangle, & \text{for } i = 1, \end{cases} \quad \text{or equivalently} \quad \boldsymbol{n} \longrightarrow \begin{cases} +\hat{z}, & \text{for } i = 0 \\ -\hat{z}, & \text{for } i = 1. \end{cases} \tag{12}$$

Without conditioning on the outcomes after the measurement the system is described by the following mixture $\rho \longrightarrow \sum_i P_{\rho}(i)|i\rangle\langle i|$ (or equivalently $\boldsymbol{n} \longrightarrow (\hat{z} \cdot \boldsymbol{n})\,\hat{z}$). These rules describe a qubit in a nutshell.

In order to be physically meaningful this framework needs to be equipped with an interpretation. Here, we will be concerned with a typical realization of a qubit associated with spatial degrees of freedom of a *single particle*, e.g. a photon traversing two arms of the Mach–Zehnder interferometer as shown in figure 3. In general, one can think of a particle which enters a system of *two paths* labelled $i = 0, 1$ arranged in a circuit built from optical gates and detectors [16, 17]. The quantum description of such a system is equivalent to a qubit where states of the computational basis, $|0\rangle$ and $|1\rangle$, correspond to the statement that the particle is respectively in the 0th or the 1st path of the circuit. The general state of the system is described by a mixed state $\rho$ (equiv. $\boldsymbol{n}$) and transformations are implemented by *phase shifters* and *beam splitters* which can be used to realize any unitary [17]. To be more specific, we have the following representation of basic interferometric gates:

| Gate | Matrix representation | Bloch representation |
|---|---|---|
| $P_0(\omega)$ | $\begin{pmatrix} \mathrm{e}^{\mathrm{i}\omega} & 0 \\ 0 & 1 \end{pmatrix}$ | $R_z(-\omega)$ |
| $P_1(\omega)$ | $\begin{pmatrix} 1 & 0 \\ 0 & \mathrm{e}^{\mathrm{i}\omega} \end{pmatrix}$ | $R_z(+\omega)$ |
| $B(\xi)$ | $\begin{pmatrix} \mathrm{i}\cos\frac{\xi}{2} & \sin\frac{\xi}{2} \\ \sin\frac{\xi}{2} & \mathrm{i}\cos\frac{\xi}{2} \end{pmatrix}$ | $R_x(\xi)$ |

where $P_i(\omega)$ denotes phase shifter placed in the $i$th path, and $B(\xi)$ stands for a beam splitter characterized by reflectivity $R = \cos^2\frac{\xi}{2}$ and transitivity $T = \sin^2\frac{\xi}{2}$. Note that phase shifters act locally on each path while beam splitters are non-local gates which require both paths brought together. Finally, a measurement in the computational basis comes down to detection of the particle whose presence in a given path is revealed by 'CLICK'/'NO CLICK' of the respective *detector* $D_i$, see equations (11) and (12).[2]

In summary, the mathematical description of a single particle in two-path interferometric circuits boils down to a qubit. We note that this framework is often used for discussion of some typically quantum effects

---

[2] Let us note that in the case of two detectors $D_0$ and $D_1$ placed in both paths of the circuit at the same time their outcomes are always anti-correlated. This leads to an interesting interpretation of the '*negative*' measurement result when there is a *single* detector, say $D_i$ in the $i$th path. Namely, given there is a particle in the system, its location can be deduced from the reading of the single detector: if it 'CLICKS' then the particle is in the $i$th path, whereas 'NO CLICK' means that the particle is in the other $\bar{i}$ th path. This observation is the basis for the analysis of interaction-free measurements; see [6–8, 16] for discussion.

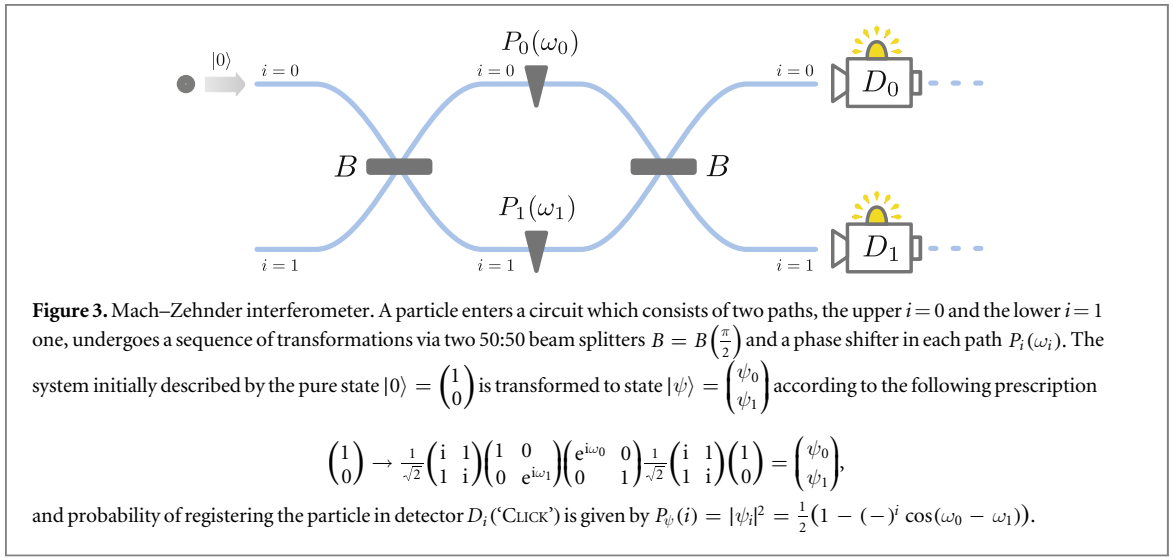

**Figure 3.** Mach–Zehnder interferometer. A particle enters a circuit which consists of two paths, the upper $i = 0$ and the lower $i = 1$ one, undergoes a sequence of transformations via two 50:50 beam splitters $B = B\left(\frac{\pi}{2}\right)$ and a phase shifter in each path $P_i(\omega_i)$. The system initially described by the pure state $|0\rangle = \begin{pmatrix} 1 \\ 0 \end{pmatrix}$ is transformed to state $|\psi\rangle = \begin{pmatrix} \psi_0 \\ \psi_1 \end{pmatrix}$ according to the following prescription

$$\begin{pmatrix} 1 \\ 0 \end{pmatrix} \rightarrow \frac{1}{\sqrt{2}} \begin{pmatrix} i & 1 \\ 1 & i \end{pmatrix} \begin{pmatrix} 1 & 0 \\ 0 & e^{i\omega_1} \end{pmatrix} \begin{pmatrix} e^{i\omega_0} & 0 \\ 0 & 1 \end{pmatrix} \frac{1}{\sqrt{2}} \begin{pmatrix} i & 1 \\ 1 & i \end{pmatrix} \begin{pmatrix} 1 \\ 0 \end{pmatrix} = \begin{pmatrix} \psi_0 \\ \psi_1 \end{pmatrix},$$

and probability of registering the particle in detector $D_i$ ('CLICK') is given by $P_\psi(i) = |\psi_i|^2 = \frac{1}{2}(1 - (-)^i \cos(\omega_0 - \omega_1))$.

[4–14] which include demonstration of the 'non-local' behaviour of quantum particles when the paths are taken to be spatially separated. In the following, we show that it is possible to simulate such interferometric setups in a classical manner using stochastic gates and two kinds of particles which interact only locally.

## 4. Ontology of the model

In this section we construct a full stochastic model of interferometric setups based on the concept of two kinds of particles propagating along the paths (see section 2). We start by defining the ontic state space of the system and discuss its probabilistic account emphasizing the distinction between local and non-local gates. Then we complete the description of the model by postulating stochastic counterparts of the interferometric gates which define the possible evolution of the system.

### 4.1. Ontic state space and stochastic gates

Let us assume that there are two kinds of particles with inner degrees of freedom: *real* ones described by a point on sphere $S^2$ in 3D, and *ghosts* described by a point on circle $S^1$ in 2D. In other words, each *real* particle carries a unit vector $\vec{n} = (\theta, \phi)$ where $\theta$ and $\phi$ are standard polar and azimuthal angles, and each *ghost* particle carries only a phase $\varphi \in [0, 2\pi)$. See figure 1 for illustration. Naming of the particles will become evident when we will introduce detectors which react only to *real* particles and remain blind to *ghosts*.

We will consider the evolution of such particles traversing a circuit which consist of *two paths* labeled $i = 0, 1$ and various types of gates which implement transformations. One may think of two *spatially separated* paths (or wires) through which particles are sent in direct analogy with quantum-interferometric circuits described in the previous section. Furthermore, we will restrict our model to situations with a *single real* particle present in one of the paths which is either accompanied by a *ghost* particle in the other path or the other path is *empty*. See figure 4 for illustration. Hence, a formal definition of the system boils down to the following description.

**Definition 1 (Ontic state pace).** At each time the system is described by a point in the ontic state space

$$\Omega \equiv \{0, 1\} \times S^2 \times S^{1*}, \tag{13}$$

where $S^{1*} \equiv S^1 \cup \{\varnothing\}$. The ontic state $(i, \vec{n}, \varphi) \in \Omega$ corresponds to a situation in which there is a *real* particle in path $i$ carrying vector $\vec{n}$ and the *ghost* particle is present in the other path $\bar{i} \equiv i + 1 \pmod{2}$ carrying phase $\varphi$. The case $(i, \vec{n}, \varnothing) \in \Omega$ describes a situation in which the (other) $\bar{i}$ th path is *empty*.

In the following we will be interested in the stochastic evolution of the system. This means that we admit incomplete knowledge which, in general, can be specified by a probability distribution over ontic states of the system.

**Definition 2 (Probabilistic description).** At each time the system is described by a point in the probabilistic simplex

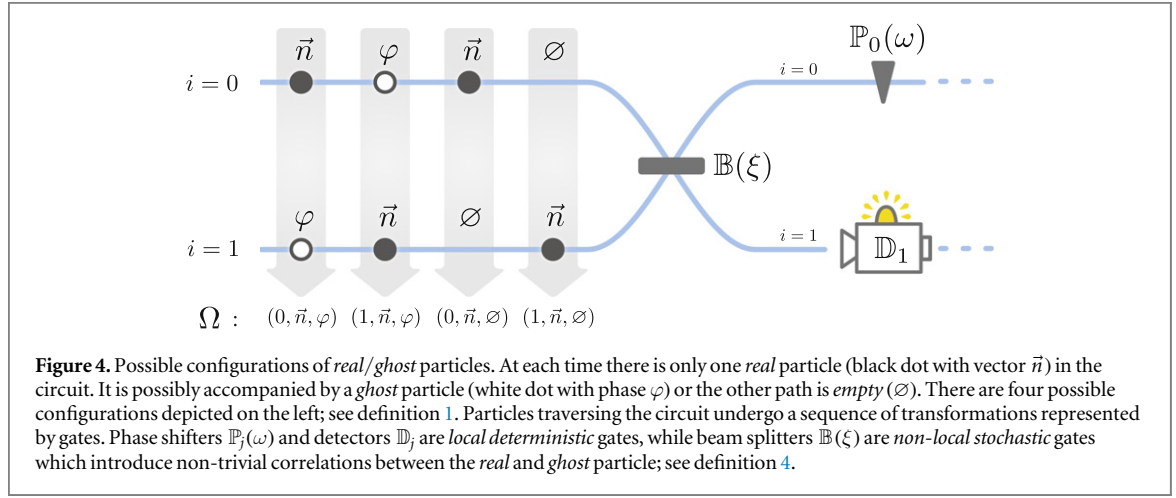

**Figure 4.** Possible configurations of *real/ghost* particles. At each time there is only one *real* particle (black dot with vector $\vec{n}$) in the circuit. It is possibly accompanied by a *ghost* particle (white dot with phase $\varphi$) or the other path is *empty* ($\varnothing$). There are four possible configurations depicted on the left; see definition 1. Particles traversing the circuit undergo a sequence of transformations represented by gates. Phase shifters $\mathbb{P}_j(\omega)$ and detectors $\mathbb{D}_j$ are *local deterministic* gates, while beam splitters $\mathbb{B}(\xi)$ are *non-local stochastic* gates which introduce non-trivial correlations between the *real* and *ghost* particle; see definition 4.

$$\mathcal{P}(\Omega) = \mathcal{P}(\{0, 1\}) \otimes \mathcal{P}(S^2) \otimes \mathcal{P}(S^{1*}), \tag{14}$$

whose elements $\boldsymbol{p} \in \mathcal{P}(\Omega)$ are probability distributions over the ontic state space $\Omega$, i.e. $\boldsymbol{p} : \Omega \longrightarrow [0, 1]$ satisfying the normalization condition $\int_\Omega \boldsymbol{p}(\omega)\mathrm{d}\omega$.

Note that ontic states $\omega = (i, \vec{n}, \tau) \in \Omega$, with $\tau = \varphi$ or $\varnothing$, correspond to extremal points of the simplex which are definite probability distributions $\delta_i\, \delta_{\vec{n}}\, \delta_\tau : \Omega \longrightarrow [0, 1]$ understood as a (tensor) product of delta functions $\delta_i\, \delta_{\vec{n}}\, \delta_\tau \equiv \delta_i \otimes \delta_{\vec{n}} \otimes \delta_\tau$, i.e. $\delta_i\, \delta_{\vec{n}}\, \delta_\tau (i', \vec{n}', \tau') \equiv \delta_{i,i'}\, \delta(\vec{n} - \vec{n}')\delta(\tau - \tau')$. They form a basis in $\mathcal{P}(\Omega)$ in the sense that each probability distribution $\boldsymbol{p} : \Omega \longrightarrow [0, 1]$ can be uniquely written in the form

$$\boldsymbol{p}(i, \vec{n}, \tau) = \sum_{i'=0,1} \int_{S^2}\!\mathrm{d}\vec{n}' \int_{S^{1*}}\!\mathrm{d}\tau'\, \boldsymbol{p}(i', \vec{n}', \tau')\delta_{i'}\, \delta_{\vec{n}'}\, \delta_{\tau'}.$$

A general *stochastic transformation* (or *gate*) is defined as a mapping

$$\mathbb{T} : \Omega \longrightarrow \mathcal{P}(\Omega), \tag{15}$$

which describes the probability distribution $\mathbb{T}(\omega) \in \mathcal{P}(\Omega)$ of the final states given the system was in the ontic state $\omega \in \Omega$. This means that $\mathbb{T}_{\omega, \omega'} \equiv \mathbb{T}(\omega)(\omega')$ is the transition probability from state $\omega$ to $\omega'$. In the following, we assume Markov property which entails that a sequence of transformations $\mathbb{T}_1, \mathbb{T}_2, \ldots, \mathbb{T}_n$ defines a new transformation associated with the product of matrices $\mathbb{T} = \mathbb{T}_1\, \mathbb{T}_2 \ldots \mathbb{T}_n$.

Note that the definition of the stochastic gate in equation (15) is fully consistent with the assertion that at each time the system is in a well-defined state. It is only our incomplete knowledge about the actual ontic state (e.g. due to random disturbance of the system by transformations) which manifests by random results at each run of the experiment. Therefore, we may think of $\boldsymbol{p} \in \mathcal{P}(\Omega)$ as describing an ensemble of systems whose elements are all in definite ontic states which are distributed according to $\boldsymbol{p}$.[3]

### 4.2. Paradigm of locality

Our primary concern in building the model is to preserve the paradigm of locality which requires that particles (the *real* ones as well as the *ghosts*) interact only with objects in their immediate vicinity. This means that in a situation of spatial separation the local gate affects *only* the inner state of the particle present in the given path and does *not* depend on the inner state of the other particle *nor* the gate implemented in the other path. For the purpose of the present model it is enough to specify the locality condition only for the case of *deterministic gates*, i.e. such that $\mathbb{T}(\omega) = \delta_{\omega'} = \delta_{i'}\, \delta_{\vec{n}'}\, \delta_{\tau'}$ with $\omega' = \omega'(\omega)$ for each $\omega \in \Omega$. We make the following formal definition.

**Definition 3 (Locality criterion).** The deterministic gate $\mathbb{T}_j$ acts *locally* in the $j$th path if and only if it can be specified by two functions

$$T_j^{(\mathrm{r})} : S^2 \longrightarrow S^2 \quad \text{and} \quad T_j^{(\mathrm{g})} : S^{1*} \longrightarrow S^{1*},$$

which define the action of the gate on the given kind of particle (*real* or *ghost/empty* respectively) present in the $j$th path and leave the other path unaffected. That is we have

$$(i, \vec{n}, \tau) \xrightarrow{\mathbb{T}_j} \begin{cases} \delta_i \, \delta_{T_j^{(r)}\vec{n}} \, \delta_\tau & \text{if } i = j, \\ \delta_i \, \delta_{\vec{n}} \, \delta_{T_j^{(g)}\tau} & \text{if } i \neq j, \end{cases} \tag{16}$$

where the case $i = j$ corresponds to the *real* particle being in the $j$th path, while $i \neq j$ means that the $j$th path contains the *ghost* or it is *empty*.

---

[3] One should contrast stochastic transformations of equation (15) with mappings of the type $\mathbb{T} : \mathcal{P}(\Omega) \longrightarrow \mathcal{P}(\Omega)$ which transform entire probability distributions. We note that in the latter case an account in terms of well-defined ontic states is no longer tenable. Since in this paper we seek explanations in reference to the underlying ontological picture, we stick to the definition of equation (15).

A general deterministic gate $\mathbb{T}$ is said to be *local* if it can be decomposed into two local deterministic gates $\mathbb{T}_j$ acting in the respective paths $j = 0, 1$, i.e. $\mathbb{T} = \mathbb{T}_0 \, \mathbb{T}_1 = \mathbb{T}_1 \, \mathbb{T}_0$. Otherwise the transformation is said to be *non-local*.

If we consider discrete time intervals, we may decompose the evolution of the system and trace the character of the consecutive steps. Usually it is presented in the graphical form of a circuit with spatially separated paths represented by lines and transformations depicted by blocks (or pictures). If the gate is known to be local then it is depicted by separate blocks attached to the respective lines, otherwise the lines 'meet' in the block indicating that particles are allowed to interact. Note that this is exactly the picture that we have for quantum-interferometric circuits discussed in section 3. It suggests an interpretation of phase shifters $P_j(\omega)$ and detectors $D_j$ as local gates, and beam splitters $B(\xi)$ being the only gates that are non-local; see figure 3. In constructing our model we will pay special attention to get identical structure of circuits, i.e. the same character of locality/non-locality for the respective gates; see figure 4.

### 4.3. Building blocks of the model

Now we are in a position to complete the model by specifying stochastic counterparts of the basic interferometric building blocks (see section 3).

**Definition 4 (Limited set of stochastic gates).** We will consider stochastic circuits that are built from a few basic building blocks defined as follows.

(i) The *phase shifter* $\mathbb{P}_j(\omega)$ is a local deterministic gate whose action is given by

$$(i, \vec{n}, \varphi) \xrightarrow{\mathbb{P}_j(\omega)} \begin{cases} \delta_i \, \delta_{R_z((-)^i\omega)\vec{n}} \, \delta_\varphi & \text{if } i = j, \\ \delta_i \, \delta_{\vec{n}} \, \delta_{\varphi+(-)^i\omega} & \text{if } i \neq j, \end{cases}$$

and

$$(i, \vec{n}, \varnothing) \xrightarrow{\mathbb{P}_j(\omega)} \begin{cases} \delta_i \, \delta_{R_z((-)^i\omega)\vec{n}} \, \delta_\varnothing & \text{if } i = j, \\ \delta_i \, \delta_{\vec{n}} \, \delta_\varnothing & \text{if } i \neq j. \end{cases}$$

Note that the phase shifter $\mathbb{P}_j(\omega)$ affects particles only in the $j$th path and the inner states of *real* and *ghost* particles are rotated around the $\hat{z}$-axis in opposite directions. Clearly, if the path is empty, it remains so.

(ii) The *beam splitter* $\mathbb{B}(\xi)$ is a non-local stochastic gate which requires information from both paths to effect the transformation. It is defined as follows

$$(i, \vec{n}, \varphi) \xrightarrow{\mathbb{B}(\xi)} \cos^2\left(\tfrac{\theta'}{2}\right) \delta_i \, \delta_{\vec{n}'} \, \delta_0 \; + \; \sin^2\left(\tfrac{\theta'}{2}\right) \delta_{\bar{\imath}} \, \delta_{-\vec{n}'} \, \delta_0,$$

where $\vec{n}' = (\theta', \phi') = R_x(\xi)R_z(-\varphi)\vec{n}$, and

$$(i, \vec{n}, \varnothing) \xrightarrow{\mathbb{B}(\xi)} \cos^2\left(\tfrac{\theta'}{2}\right) \delta_i \, \delta_{\vec{n}'} \, \delta_0 \; + \; \sin^2\left(\tfrac{\theta'}{2}\right) \delta_{\bar{\imath}} \, \delta_{-\vec{n}'} \, \delta_0,$$

where $\vec{n}' = (\theta', \phi') = R_x(\xi)\hat{z}$.

In the first case, the resulting state is a probabilistic mixture of two situations: particles remain in their respective paths ($i \rightarrow i$) or particles get swapped ($i \rightarrow \bar{\imath}$). Note that inner states of the particles change, i.e. $\vec{n} \rightarrow \pm\vec{n}'$ and $\varphi \rightarrow 0$ with $\vec{n}'$ depending on $\vec{n}$ and $\varphi$, and the combination of terms in the mixture features non-trivial correlations. In the second case, the action of the beam splitter can be described in a similar way, but first the lacking information in the *empty* path ($\varnothing$) needs to be completed by creating there a *ghost* with $\varphi = 0$ and changing $\vec{n} \rightarrow \hat{z}$ for the *real* particle.

(iii) The *detector* $\mathbb{D}_j$ is a local deterministic gate which 'Clicks' if it finds a *real* particle in the $j$th path and remains silent otherwise ('No Click'). It has the following action

$$(i, \vec{n}, \varphi) \xrightarrow{\mathbb{D}_j} \begin{cases} \delta_i \, \delta_{\hat{z}} \, \delta_\varphi & \text{if } i = j \text{ ('Click'),} \\ \delta_i \, \delta_{\vec{n}} \, \delta_\varnothing & \text{if } i \neq j \text{ ('No Click'),} \end{cases}$$

and

$$(i, \vec{n}, \varnothing) \xrightarrow{\ \mathbb{D}_j\ } \begin{cases} \delta_i \ \delta_{\hat{z}} \ \delta_{\varnothing} & \text{if} \quad i = j \quad (\text{`Click'}), \\ \delta_i \ \delta_{\vec{n}} \ \delta_{\varnothing} & \text{if} \quad i \neq j \quad (\text{`No Click'}). \end{cases}$$

Note that the detector $\mathbb{D}_j$ acts nontrivially on particles of either type. If it happens to be a *real* particle ($i = j$) then $\vec{n} \to \hat{z}$. If it is a *ghost* ($i \neq j$) then it gets absorbed and the path is left *empty*, i.e. $\varphi \to \varnothing$. Clearly, detection is repeatable (i.e. subsequent detection gives the same result).

Notice that definition of the detection process via $\mathbb{D}_j$'s explains the naming of the particles: the *real* ones are those 'observed' by detectors ('Click'), while the *ghosts* are particles invisible to detectors ('No Click').

It is instructive to remark that the beam splitter $\mathbb{B}(\xi)$ is the only place in the model where the *ghosts* can manifest their presence. When looking more closely at the definition of $\vec{n}'$, it appears that phase the *ghost* particle provides a kind of 'reference frame' for the transformation effected in the beam splitter. One can think of this as information transfer between particles in different paths (it is allowed since the beam splitter is a non-local gate where both paths meet). If not this effect, predictions of the model would be insensitive to existence of the *ghosts*. In the following we will show that their presence leads to 'observable' consequences in circuits composed of a sequence of gates.

Clearly, we can replace quantum gates with their stochastic counterparts to imitate any two-path interferometric circuit (see section 3). So far this is the only structural similarity and it is not clear why the analogy should go further. Although it might come as surprise in view of the fact that the model is built on the classical-like particle ontology with both kinds of particles as well the action of the gates conforming to the paradigm of locality, we show in the following section that the operational description of the model is equivalent to a qubit.

## 5. Operational description of the model

Imagine an agent without any prior notion of the model trying to make sense of how it works only on the basis of experiments she performs. Clearly, the agent acts under *epistemic constraints* which confine her perception of the system under study—she has a limited choice of gates for building circuits and her detectors 'see' only *real* particles (see definition 4). That being the case the agent describing the model is legitimate in restricting herself only to situations within her reach. In other words, for all practical purposes it is sufficient to account for a limited set of preparation, transformation and measurement procedures that arise in any experimental circuit that she can build according to the rules of the model. Note that from this point of view any ontological commitment is superfluous and a minimal mathematical framework is just enough.

In the following, the full-blown ontological picture of the model is cut down to a minimal *operational account* adapted for the specific needs of the agent. We proceed in steps by answering the following questions.

(a) Which distributions in $\mathcal{P}(\Omega)$ can be prepared by the agent according the rules of the model?

(b) How do they transform and what information can be learned under the action of conceivable circuits?

We will see that the agent has access only to a limited range of distributions which form a well-structured set under possible transformations (section 5.1). A closer look will reveal that a detailed ontological account is in many respects irrelevant, which brings up another question:

(c) What is the minimal description which is enough to predict behaviour of the system?

Our main result shows that the operational account of the system is equivalent to a qubit (section 5.2). It means that from the agent's perspective predictions of the model are indistinguishable from the behaviour of a single quantum particle in two-path interferometric circuits discussed in section 3.

### 5.1. Ontological account
#### 5.1.1. Initializing the circuit
Before any analysis takes place we need to address the problem of initialization, i.e. how the agent starts the circuit off by preparing a reliable ensemble of particles for which it is going to work. We know from section 4.1 that in order to meet the requirements posed by the model it has to be an ensemble of particles with the property that at each time there is only a single *real* particle present in one of the paths and possibly a *ghost* in the other one, i.e. an ensemble described by a distribution in $\mathcal{P}(\Omega)$.

It must be realized that without access to any particular source from the outside the agent has to find a method to prepare the initial ensemble only by herself. The least that can be assumed is that the agent is given access to an unknown (possibly random) source of particles. Then, to make it a reliable initial ensemble it has to be sieved in search of cases in which the circuit will certainly work by checking for the presence of *real* particles in the paths. This can be done by filtering via detectors placed in both paths, $\mathbb{D}_0$ and $\mathbb{D}_1$, and retaining only the cases when a single detection occurred.[4] See figure 5 for illustration. It secures proper working of the circuit and provides the least framework to start investigation of the model. Thus, upon selection of events when a given detector $\mathbb{D}_i$ 'CLICKED' the agent prepares two initial ensembles described by the distributions (see definition 4 (iii))

$$\boldsymbol{p}_i^{(\mathrm{in})} \equiv \delta_i \, \delta_{\hat{z}} \, \delta_{\varnothing} \in \mathcal{P}(\Omega). \tag{17}$$

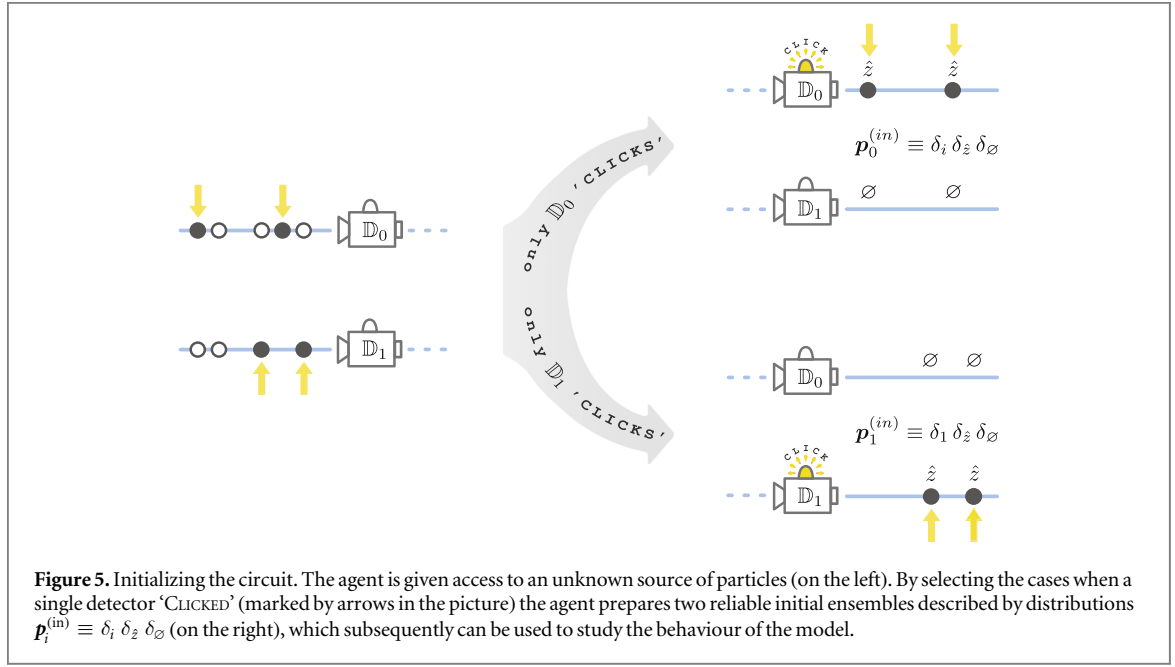

**Figure 5.** Initializing the circuit. The agent is given access to an unknown source of particles (on the left). By selecting the cases when a single detector 'CLICKED' (marked by arrows in the picture) the agent prepares two reliable initial ensembles described by distributions $\boldsymbol{p}_i^{(\mathrm{in})} \equiv \delta_i \, \delta_{\hat{z}} \, \delta_{\varnothing}$ (on the right), which subsequently can be used to study the behaviour of the model.

From the ontic point of view the states in equation (17) correspond to the *real* particle with inner state $\vec{n} = \hat{z}$ being present in the $i$th path and no *ghost* in the another path. We note in advance that such a detailed description is not accessible to the agent. The only knowledge that she has is that the $i$th detector 'CLICKED' and immediate repetition of the test will necessarily yield the same outcome.

Certainly, the agent can go beyond the initial states in equation (17) by processing them with the tools that she has at her disposal (phase shifters, beam splitters and detectors). Combining gates and measurements in various orders broadens the scope of the preparation procedures leading to a considerable variety of distributions which will be systematically characterized in the following two subsections.

### 5.1.2. Some states of interest and action of the gates

For the purpose of analysis let us distinguish a few important classes of distributions in $\mathcal{P}(\Omega)$. We will denote them by $[\vec{N}]$ and label by unit vectors $\vec{N} = (\theta, \phi) \in \mathcal{S}^2$, where $\theta$ and $\phi$ are standard polar and azimuthal angles. See figure 6 for illustration.

For each $\vec{N} \neq \pm\hat{z}$ we define the corresponding class $[\vec{N}] \subset \mathcal{P}(\Omega)$ as follows

$$[\vec{N}] \equiv \left\{ \boldsymbol{p}_{\vec{N}}^{(\alpha, \beta)} : \; \alpha, \beta \in [0, 2\pi) \right\}, \tag{18}$$

where

$$\boldsymbol{p}_{\vec{N}}^{(\alpha, \beta)} \equiv \cos^2\left(\frac{\theta}{2}\right) \, \delta_0 \, \delta_{R_{\hat{z}}(\alpha)\vec{N}} \, \delta_\alpha \; + \; \sin^2\left(\frac{\theta}{2}\right) \, \delta_1 \, \delta_{-R_{\hat{z}}(\beta)\vec{N}} \, \delta_\beta. \tag{19}$$

All these distributions correspond to both *real* and *ghost* particles present in the system. Moreover, each distribution $\boldsymbol{p}_{\vec{N}}^{(\alpha, \beta)} \in [\vec{N}]$ is a mixture of two definite situations: with the *real* particle being present either in the upper path ($i = 0$) or in the lower path ($i = 1$), with the respective probabilities $P_{\vec{N}}(i = 0) = \cos^2\left(\frac{\theta}{2}\right)$ and $P_{\vec{N}}(i = 1) = \sin^2\left(\frac{\theta}{2}\right)$ which depend only on the polar angle $\theta$ of the defining vector $\vec{N} = (\theta, \phi)$. Notice that in the distributions $\boldsymbol{p}_{\vec{N}}^{(\alpha, \beta)}$ defined in equation (19) the inner states of both particles are closely related, i.e. the vector carried by the *real* particle is given by $\pm \vec{N}$ (with the sign depending on where it is $i = 0, 1$) rotated about the $\hat{z}$-axis by an angle equal to the phase of the *ghost* ($\alpha$ or $\beta$ respectively). It is important to realize that distributions defined in this way display correlations, which will play a crucial role in the following analysis.[5]

---

[4] Alternatively, one can use a single detector and block the other path, but this gives the same initial distributions as in equation (17).

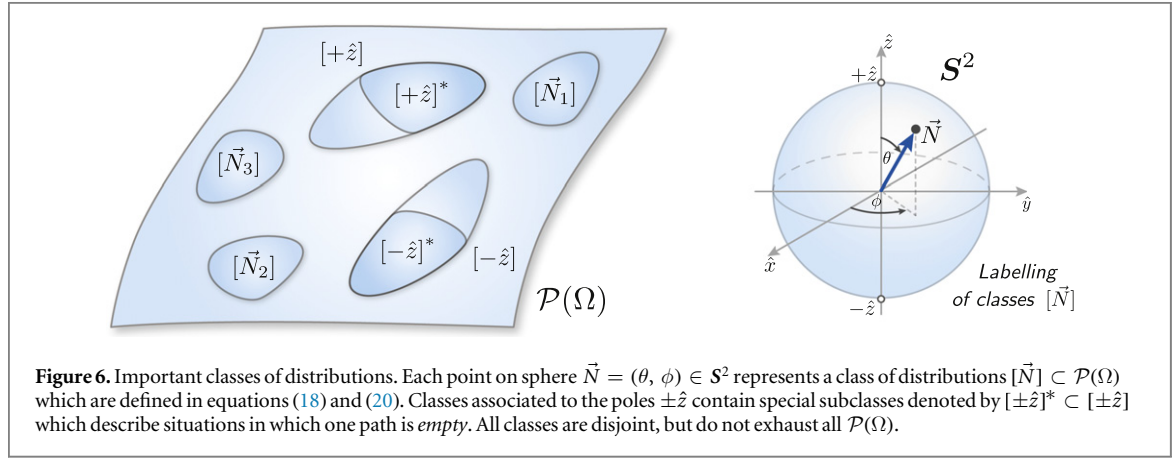

**Figure 6.** Important classes of distributions. Each point on sphere $\vec{N} = (\theta, \phi) \in S^2$ represents a class of distributions $[\vec{N}] \subset \mathcal{P}(\Omega)$ which are defined in equations (18) and (20). Classes associated to the poles $\pm \hat{z}$ contain special subclasses denoted by $[\pm \hat{z}]^* \subset [\pm \hat{z}]$ which describe situations in which one path is *empty*. All classes are disjoint, but do not exhaust all $\mathcal{P}(\Omega)$.

For the north and south pole $\vec{N} = \pm \hat{z}$, we make the following definitions

$$[+\hat{z}] \equiv \Big\{ \delta_0 \, \delta_{\hat{z}} \, \delta_\alpha : \ \alpha \in [0, 2\pi) \Big\} \ \cup \ [+\hat{z}]^*,$$
$$[-\hat{z}] \equiv \Big\{ \delta_1 \, \delta_{\hat{z}} \, \delta_\beta : \ \beta \in [0, 2\pi) \Big\} \ \cup \ [-\hat{z}]^*, \qquad (20)$$

where

$$[+\hat{z}]^* \equiv \Big\{ \delta_0 \, \delta_{\vec{n}} \, \delta_\varnothing : \vec{n} \in S^2, \, \vec{n} \neq -\hat{z} \Big\},$$
$$[-\hat{z}]^* \equiv \Big\{ \delta_1 \, \delta_{\vec{n}} \, \delta_\varnothing : \vec{n} \in S^2, \, \vec{n} \neq -\hat{z} \Big\}. \qquad (21)$$

Every distribution in the class $[+\hat{z}]$ (resp. $[-\hat{z}]$) corresponds to a situation with the *real* particle being definitely in the 0th (resp. 1st) path of the circuit. Note that each class $[\pm \hat{z}]$ is composed of two subclasses distinguished by the presence or absence of the *ghost*. In the case when the *ghost* is present in the system, the inner state of the *real* particle always points north $\vec{n} = \hat{z}$. Otherwise, when the other path is *empty* ($\varnothing$), the vector carried by the *real* particle can point in any direction with the exception of the south pole $\vec{n} \in S^2 \backslash \{-\hat{z}\}$. We denote the latter subclasses by $[\pm \hat{z}]^*$. Clearly, for the initial distributions in equation (17), we have $\boldsymbol{p}_0^{(\mathrm{in})} \in [+\hat{z}]^*$ and $\boldsymbol{p}_1^{(\mathrm{in})} \in [-\hat{z}]^*$.

In summary, we have associated to each point on sphere $S^2$ the corresponding class of distributions in $\mathcal{P}(\Omega)$, i.e. we have a mapping $S^2 \ni \vec{N} \longrightarrow [\vec{N}] \subset \mathcal{P}(\Omega)$. See figure 6 for illustration. One readily checks that these classes are disjoint, i.e.

$$[\vec{N}] \cap [\vec{N}'] = \varnothing \quad \text{for} \quad \vec{N} \neq \vec{N}'. \qquad (22)$$

However, it is not a partition of $\mathcal{P}(\Omega)$ since they do not exhaust all possible distributions in $\mathcal{P}(\Omega)$, i.e.

$$\mathcal{E} \equiv \bigcup_{\vec{N} \in S^2} [\vec{N}] \subsetneq \mathcal{P}(\Omega). \qquad (23)$$

In such a case, we say that the collection of all classes $\{[\vec{N}] : \vec{N} \in S^2\}$ defines a *partial equivalence relation* on $\mathcal{P}(\Omega)$. Clearly, in the restricted domain $\mathcal{E}$ it is a full-blown partition which defines an equivalence relation on $\mathcal{E}$ with two distributions being equivalent when they belong to the same class.

Now, let us focus on the restricted set of distributions $\mathcal{E} \subset \mathcal{P}(\Omega)$ and characterize the action of the stochastic gates defined in section 4.3. We have the following two lemmas (see appendix A for the proofs).

**Lemma 1 (Phase shifters and beam splitters).** *The action of the phase shifters* $\mathbb{P}_j(\omega)$ *and beam splitters* $\mathbb{B}(\xi)$ *does not leave outside the set* $\mathcal{E}$*, i.e. we have*

$$\mathbb{T}|_{\mathcal{E}} : \mathcal{E} \longrightarrow \mathcal{E}, \qquad (24)$$

*for any* $\mathbb{T} = \mathbb{P}_j(\omega)$ *or* $\mathbb{B}(\xi)$. *Moreover, all elements of a given class are mapped (congruently) into the same class, i.e.* $[\vec{N}] \ni \boldsymbol{p} \overset{\mathbb{T}}{\longrightarrow} \boldsymbol{p}' \in [\vec{N}_{\mathbb{T}}]$ *with* $\vec{N}_{\mathbb{T}}$ *depending only on* $\vec{N}$ *and the transformation* $\mathbb{T}$.

---

[5] Distributions $\boldsymbol{p}_{\vec{N}}^{(\alpha, \beta)}$ do not factorize in variables $i$, $\vec{n}$ and $\varphi$. Note also the curious correlation between the position of the *real* particle $i = 0$, 1 and the $\pm$ sign of its inner state $+R_z(\alpha)\vec{N}$ or $-R_z(\beta)\vec{N}$ relative to the phase of the *ghost* ($\alpha$ or $\beta$ respectively).

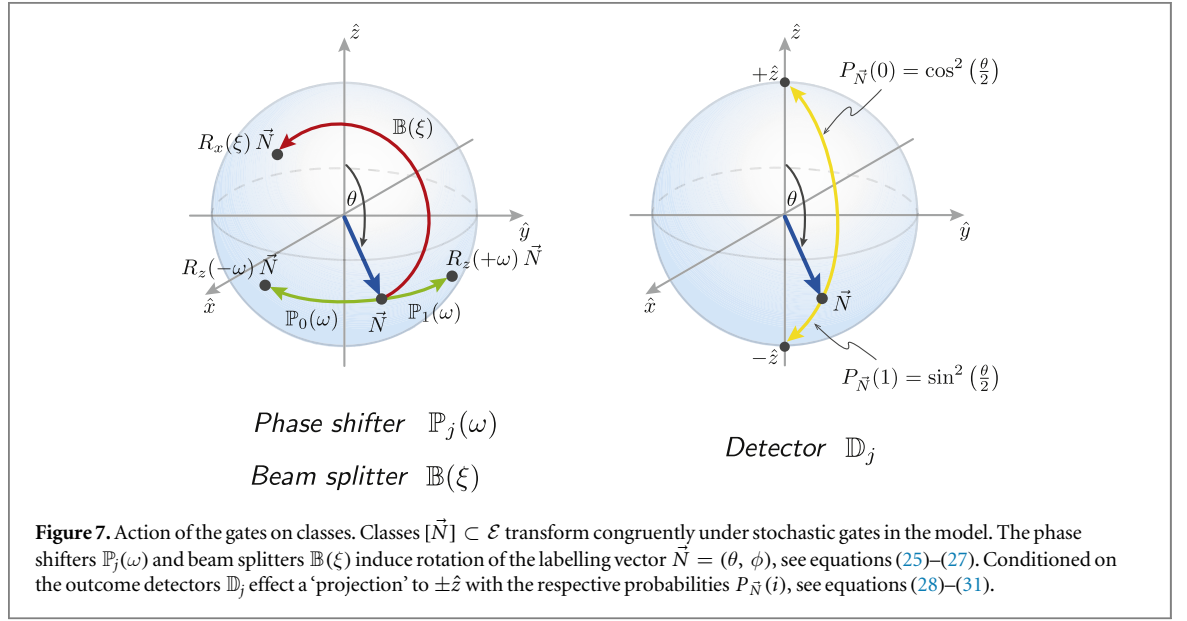

**Figure 7.** Action of the gates on classes. Classes $[\vec{N}] \subset \mathcal{E}$ transform congruently under stochastic gates in the model. The phase shifters $\mathbb{P}_j(\omega)$ and beam splitters $\mathbb{B}(\xi)$ induce rotation of the labelling vector $\vec{N} = (\theta, \phi)$, see equations (25)–(27). Conditioned on the outcome detectors $\mathbb{D}_j$ effect a 'projection' to $\pm \hat{z}$ with the respective probabilities $P_{\vec{N}}(i)$, see equations (28)–(31).

*For phase shifters* $\mathbb{P}_j(\omega)$, *we have the following rules*

$$[\vec{N}] \ni \boldsymbol{p} \xrightarrow{\mathbb{P}_0(\omega)} \boldsymbol{p}' \in [R_z(-\omega)\vec{N}], \tag{25}$$

$$[\vec{N}] \ni \boldsymbol{p} \xrightarrow{\mathbb{P}_1(\omega)} \boldsymbol{p}' \in [R_z(+\omega)\vec{N}]. \tag{26}$$

*For beam splitters* $\mathbb{B}(\xi)$, *we get*

$$[\vec{N}] \ni \boldsymbol{p} \xrightarrow{\mathbb{B}(\xi)} \boldsymbol{p}' \in [R_x(\xi)\vec{N}]. \tag{27}$$

In other words, classes of distributions $[\vec{N}] \subset \mathcal{E}$ transform as a whole (congruently) under the action of phase shifters and beam splitters, and the mapping of classes boils down to appropriate rotation of the labelling vector $\vec{N} \longrightarrow R_F(\gamma)\vec{N}$ as specified in equations (25)–(27). See figure 7 (left) for illustration.

In the case of detectors the situation is subtler since they provide additional information given by the outcome, i.e. 'CLICK' or 'NO CLICK', which should be included in the description. In general, there are three different arrangements: with a single detector detector $\mathbb{D}_j$ in one of the paths ($j = 0, 1$) or two detectors $\mathbb{D}_0$ and $\mathbb{D}_1$ placed in both paths at the same time. We get the following description.

**Lemma 2** (**Detectors**). *Suppose that we have an ensemble which is described by the distribution* $\boldsymbol{p} \in [\vec{N}]$ *with* $\vec{N} = (\theta, \phi)$, *and there is a single detector* $\mathbb{D}_j$ *placed in the jth paths.*
*For the detector* $\mathbb{D}_0$, *its action is given by*

– $\mathbb{D}_0$'*Clicks' with probability* $P_{\vec{N}}(i = 0) = \cos^2\left(\frac{\theta}{2}\right) = \frac{1}{2}(1 + \hat{z} \cdot \vec{N})$,

– $\mathbb{D}_0$ *does not 'Click' with probability* $P_{\vec{N}}(i = 1) = \sin^2\left(\frac{\theta}{2}\right) = \frac{1}{2}(1 - \hat{z} \cdot \vec{N})$, $\tag{28}$

*and afterwards the system is left in state described by the distribution*

$$[\vec{N}] \ni \boldsymbol{p} \xrightarrow{\mathbb{D}_0} \begin{cases} \boldsymbol{p}' \in [+\hat{z}] & \text{if } \mathbb{D}_0\text{'CLICKED'}, \\ \boldsymbol{p}' \in [-\hat{z}] & \text{if } \mathbb{D}_0 \text{ did not 'CLICK'}. \end{cases} \tag{29}$$

*For the detector* $\mathbb{D}_1$, *we have*

– $\mathbb{D}_1$'*Clicks' with probability* $P_{\vec{N}}(i = 1) = \sin^2\left(\frac{\theta}{2}\right) = \frac{1}{2}(1 - \hat{z} \cdot \vec{N})$,

– $\mathbb{D}_1$ *does not 'Click' with probability* $P_{\vec{N}}(i = 0) = \cos^2\left(\frac{\theta}{2}\right) = \frac{1}{2}(1 + \hat{z} \cdot \vec{N})$, $\tag{30}$ and afterwards the system is left in state described by the distribution

$$[\vec{N}] \ni \boldsymbol{p} \xrightarrow{\mathbb{D}_1} \begin{cases} \boldsymbol{p}' \in [-\hat{z}] & \text{if } \mathbb{D}_1 \text{`CLICKED'}, \\ \boldsymbol{p}' \in [+\hat{z}] & \text{if } \mathbb{D}_1 \text{ did not `CLICK'}. \end{cases} \tag{31}$$

In the case of two detectors $\mathbb{D}_0$ and $\mathbb{D}_1$ placed in both paths at the same time their 'CLICKS' are anti-correlated, but otherwise their behaviour follows the rules given above.

This means that on the level of classes detection comes down to 'projection' of the labelling vector $\vec{N} \longrightarrow \pm\hat{z}$ which depends on the outcome as specified in equations (29) and (31). See figure 7 (right) for illustration. We note that without conditioning detection leads outside $\mathcal{E}$, i.e. the whole ensemble is a mixture of two distributions in $[\pm\hat{z}]$. This aspect will be discussed in the following subsection.

It is important to remark that the mappings in equations (25)–(31) act nontrivially within each class, i.e. the resulting $\boldsymbol{p}' \in [\vec{N}]$ depends on the input $\boldsymbol{p} \in [\vec{N}]$. We have skipped these details in the above statements since they are not essential for the following discussion. For explicit action of the gates see appendix A, where both lemmas 1 and 2 are proved.

### 5.1.3. Complete description of accessible states

We have discussed above a restricted set of distributions $\mathcal{E} \subset \mathcal{P}(\Omega)$ and its partition into classes $[\vec{N}] \subset \mathcal{E}$ labelled by vectors $\vec{N} \in \mathcal{S}^2$. Now, we will use them to characterize the full range of states accessible to the agent who acts in accordance with the rules of the model. This means that we have to account for all distributions obtained from the initial ensemble in any kind of procedure that is allowed by the model, i.e.

(a) any sequence of gates (circuit),

(b) possible conditioning on the outcomes,

(c) probabilistic mixing of ensembles.

In the following we argue that the agent can not reach outside the set of distributions $\mathcal{E} \subset \mathcal{P}(\Omega)$ and probabilistic mixtures thereof.

Let us start by considering an ensemble described by one of the initial distributions $\boldsymbol{p}_i^{(\text{in})}$ defined in equation (17). Clearly, we have

$$\boldsymbol{p}_0^{(\text{in})} \in [+\hat{z}] \quad \text{and} \quad \boldsymbol{p}_1^{(\text{in})} \in [-\hat{z}]. \tag{32}$$

Then lemma 1 guarantees that after processing through a circuit composed of a sequence of phase shifters and beam splitters the system ends in a state described by distribution within the restricted set $\mathcal{E}$. Moreover, equations (25)–(27) specify in which class $[\vec{N}]$ the resulting distribution will be contained—it is specified by a sequence of rotations corresponding to the gates in the circuit acting on the vector labelling the respective initial class $[\pm\hat{z}]$. For example, if the circuit consists of a sequence of gates $\mathbb{T}_1, \mathbb{T}_2, \ldots, \mathbb{T}_n$, then we have

$$[\pm\hat{z}] \longrightarrow [\pm R_{\hat{r}_n}(\gamma_n)\ldots R_{\hat{r}_2}(\gamma_2)R_{\hat{r}_1}(\gamma_1)\hat{z}], \tag{33}$$

where $R_{\hat{r}_l}(\gamma_l)$ are rotations about axes $\hat{n} = \hat{z}$ or $\hat{x}$ corresponding to the respective gates $\mathbb{T}_l$ (see lemma 1).

If there are detectors placed along the circuit, then during evolution each element of the ensemble is tagged with the respective outcomes registered by the detectors. It follows from lemma 2 that upon selection of cases corresponding to the same sequence of outcomes the associated subensemble is described by a well-defined distribution in the restricted set $\mathcal{E}$. To put it differently, each instance of detection which discloses position of the *real* particle, splits the original ensemble into two groups in which all elements are described by the same distribution in the respective class $[\pm\hat{z}]$, see equations (29) and (31). In terms of classes this induces outcome dependent 'projection', i.e. we have

$$[\vec{N}] \longrightarrow \begin{cases} [+\hat{z}], & \text{real particle in path } i = 0, \\ [-\hat{z}], & \text{real particle in path } i = 1, \end{cases} \tag{34}$$

which happens with relative frequencies given by the formula (see equations (28) and (30) in lemma 2)

$$P_{\vec{N}}(i) = \tfrac{1}{2}(1 + (-)^i \, \hat{z} \cdot \vec{N}). \tag{35}$$

Then the selected subensemble evolves again in accord with lemma 1 to a well-defined distribution in some other class $[\vec{N}] \subset \mathcal{E}$ as specified by equation (33) until the next detection takes place and the procedure repeats. Hence, upon conditioning on the readings of all detectors along the way the description of the selected subensemble follows a path in the restricted set of distributions $\mathcal{E} \subset \mathcal{P}(\Omega)$.

In fact, we may generalize this analysis to ensembles described by any initial state $\boldsymbol{p} \in [\vec{N}] \subset \mathcal{E}$ that the agent might start with. By the same reasoning we conclude that processing via any conceivable circuit and conditioning on the outcomes does not lead outside the restricted set $\mathcal{E} \subset \mathcal{P}(\Omega)$. Moreover, evolution of classes $[\vec{N}] \subset \mathcal{E}$ containing the corresponding distributions is given in the case of phase shifters $\mathbb{P}_j(\omega)$ and beam splitters $\mathbb{B}(\xi)$ by a sequence of appropriate rotations of the labelling vector (see equation (33))

$$[\vec{N}] \longrightarrow \left[ R_{\hat{r}}(\gamma) \vec{N} \right], \tag{36}$$

and outcome dependent 'projections' of equation (34) in the case of detector $\mathbb{D}_j$.

Let us summarize by the following observation (see appendix B for the proof of part (ii)).

**Proposition 1.**

(i) *Processing distributions form the restricted set $\mathcal{E} \subset \mathcal{P}(\Omega)$ via any sequence of gates and conditioning on the outcomes does not leave outside $\mathcal{E}$.*

(ii) *The agent can prepare ensembles described by any distribution in $\mathcal{E}$ by processing initial ensembles $\boldsymbol{p}_0^{(in)}$ of equation (17) via an appropriate sequence of gates.*

To exhaust all possibilities in which the agent can prepare new ensembles we need to include probabilistic mixing into the picture. This means that having a way to prepare ensembles described by distributions $\boldsymbol{p}_1, \ldots, \boldsymbol{p}_K \in \mathcal{E}$ the agent can also prepare probabilistic mixtures of the form[6]

$$\boldsymbol{p} = \sum_k p_k \, \boldsymbol{p}_k \in \mathcal{P}(\Omega), \tag{37}$$

where $p_k \geqslant 0$ and $\sum_k p_k = 1$. Clearly, if $\boldsymbol{p}_k \in [\vec{N}_k]$, then we have

$$\boldsymbol{p} \in \sum_k p_k \, [\vec{N}_k]. \tag{38}$$

We note that mixtures of this type arise, for example, when the agent does not condition on the detection outcomes. In such a case the entire ensemble after a measurement in state $\boldsymbol{p} \in [\vec{N}]$ is described by a distribution $\boldsymbol{p}' = P_{\vec{N}}(0)\boldsymbol{p}'_0 + P_{\vec{N}}(1)\boldsymbol{p}'_1$, with $\boldsymbol{p}'_0 \in [+\hat{z}]$ and $\boldsymbol{p}'_1 \in [-\hat{z}]$ and the respective probabilities $P_{\vec{N}}(i)$ specified in lemma 2. On the level of classes $[\vec{N}] \subset \mathcal{E}$ this means that without conditioning on the outcomes we get

$$[\vec{N}] \longrightarrow P_{\vec{N}}(0)[+\hat{z}] + P_{\vec{N}}(1)[-\hat{z}], \tag{39}$$

with $P_{\vec{N}}(i) = \frac{1}{2}(1 + (-)^i \, \hat{z} \cdot \vec{N})$; see equations (34) and (35).

From linearity of stochastic gates we deduce that processing of probabilistic mixtures via any conceivable circuit does not lead to further extension of the set of accessible distributions. In this way, having considered all available possibilities for exploring the model, we conclude that the agent remains confined within distributions being probabilistic mixtures of states in the restricted set $\mathcal{E}$. Hence the following result.

**Proposition 2.** *Most general distributions in $\mathcal{P}(\Omega)$ which are accessible to the agent are probabilistic mixtures of states in $\mathcal{E}$, i.e. the agent explores only the following set*

$$\mathsf{conv}\, \mathcal{E} = \left\{ \sum_k p_k \, \boldsymbol{p}_k : \, \boldsymbol{p}_k \in \mathcal{E} \right\} \subset \mathcal{P}(\Omega), \tag{40}$$

*where $p_k \geqslant 0$ and $\sum_k p_k = 1$, and the sums are finite.*

We observe that having discussed the action of the gates on the distributions in the restricted set $\mathcal{E}$, we can extend these results to trace evolution of ensembles described by any distribution in the set $\mathsf{conv}\, \mathcal{E}$. By virtue of linearity of the gates we get

$$\sum_k p_k \, \boldsymbol{p}_k \longrightarrow \sum_k p_k \, \boldsymbol{p}'_k, \tag{41}$$

where the transformation $\boldsymbol{p}_k \longrightarrow \boldsymbol{p}'_k$ follows the prescriptions of lemmas 1 and 2. By the same token we get an evolution of mixtures on the level of classes which follows

---

[6] One way of preparing such a probabilistic mixture is to use an external source of randomness which allows for a probabilistic choice of procedures which prepare the respective states $\boldsymbol{p}_1, \ldots, \boldsymbol{p}_K \in \mathcal{E}$. Alternatively, the agent can draw on the internal source of randomness generated by the probabilistic nature of the transformations effected by the beam splitters. For example, this can be implemented by preparing an ensemble described by a distribution in $\boldsymbol{p} \in [\vec{N}]$, and then performing measurement $\mathbb{D}_0$ and $\mathbb{D}_1$ without registering the outcomes (i.e. no selection of subensembles corresponding to the respective outcomes). The resulting ensemble is described by the distribution $P_{\vec{N}}(0)\boldsymbol{p}_0^{(in)} + P_{\vec{N}}(1)\boldsymbol{p}_1^{(in)}$. This state can be processed again and in this way by appropriate repetition of non-selective measurements any convex combination of distributions in $\mathcal{E}$ can be obtained.

$$\sum_k p_k \, [\vec{N}_k] \longrightarrow \sum_k p_k \, [\vec{N}_k'],$$ (42)

with $\vec{N}_k \longrightarrow \vec{N}_k'$ specified by equations (34), (36) and (39).

This completes the description of the model as experienced by the agent acting under epistemic constraints. Note that it is given from the position of an outside observer with full knowledge of the underlying ontology (it is the case of ourselves being acquainted with details of the model as laid out in section 4). However, it is important to realize that from the agent's perspective, who is unaware of the ontological aspects of the system, such a description is untenable. In the next section we show how to cut this ontological account down to size so that it complies with the specific needs and limitations of the agent.

### 5.2. Operational account and recovery of quantum description

Now we come back to the problem of describing the model without commitment to the underlying ontology. This is the situation of an agent trying to make sense of the system only by interacting with it through experiments. In such a case an ontic account of the epistemic constraints in the form given in section 5.1 is unsuitable and in many respects superfluous. For all practical purposes it suffices for the agent to focus on a minimal description which is just enough to properly predict the behaviour of the system only in situations that she may come up with. With this in mind we will adopt the results of the previous section 5.1 to construct a purely operational account of the model as seen by the agent being completely unaware of the underlying ontology as described in section 4.

For the purpose of analysis we start by introducing the appropriate notion of equivalence on the set of accessible distributions conv $\mathcal{E}$. Note that a general probabilistic mixture $\boldsymbol{p} \in$ conv $\mathcal{E}$ has the form (see equations (40) and (23))

$$\boldsymbol{p} = \sum_k p_k \, \boldsymbol{p}_k, \quad \text{with } \boldsymbol{p}_k \in [\vec{N}_k] \quad \text{and} \quad \vec{N}_k \in \boldsymbol{S}^2.$$ (43)

It follows that we can associate to each $\boldsymbol{p} \in$ conv $\mathcal{E}$ a vector defined as

$$\vec{N}_{\boldsymbol{p}} = \sum_k p_k \, \vec{N}_k.$$ (44)

Clearly, this vector has length $|\vec{N}_{\boldsymbol{p}}| \leqslant 1$ and lies in the unit ball $\vec{N}_{\boldsymbol{p}} \in \boldsymbol{B}^3$. Hence we get a mapping

$$\text{conv } \mathcal{E} \ni \boldsymbol{p} \longrightarrow \vec{N}_{\boldsymbol{p}} \in \boldsymbol{B}^3.$$ (45)

It is crucial to note that the definition of vector $\vec{N}_{\boldsymbol{p}}$ is unambiguous (see appendix C for the proof).

**Proposition 3.** *For each $\boldsymbol{p} \in$ conv $\mathcal{E}$ the corresponding vector $\vec{N}_{\boldsymbol{p}}$ given in equation* (44) *is defined uniquely, i.e. does not depend on a particular decomposition in equation* (43).

Clearly, there are many distributions corresponding to the same vector $\vec{N} \in \boldsymbol{B}^3$. For example, any distribution which is obtained from equation (43) by a different choice of representatives $\boldsymbol{p}_k' \in [\vec{N}_k]$, or another decomposition $\vec{N} = \sum_k p_k' \, \vec{N}_k'$ (with $\vec{N}_k' \in \boldsymbol{S}^2$ and $p_k' \geqslant 0$, $\sum_k p_k' = 1$), is associated to the same vector given in equation (44).

Now we are in position to define the equivalence relation on conv $\mathcal{E}$. We say that two distributions $\boldsymbol{p}, \boldsymbol{q} \in$ conv $\mathcal{E}$ are equivalent if

$$\boldsymbol{p} \sim \boldsymbol{q} \Leftrightarrow \vec{N}_{\boldsymbol{p}} = \vec{N}_{\boldsymbol{q}}.$$ (46)

It is straightforward to see that each $\vec{N} \in \boldsymbol{B}^3$ defines an equivalence class of distributions in conv $\mathcal{E}$ which has the form[7]

$$[[\vec{N}]] \equiv \left\{ \, \boldsymbol{p} \in \text{conv } \mathcal{E} : \ \vec{N}_{\boldsymbol{p}} = \vec{N} \, \right\}.$$ (47)

Clearly, these classes are disjoint, i.e.

$$[[\vec{N}]] \cap [[\vec{N}']] = \varnothing \ \text{for} \ \vec{N} \neq \vec{N}',$$ (48)

and define a partition of the set conv $\mathcal{E}$, i.e.

$$\bigcup_{\vec{N} \in \boldsymbol{B}^3} [[\vec{N}]] = \text{conv } \mathcal{E}.$$ (49)

It is important to realize that the agent collects information about the system only via measurements (i.e. 'CLICKS'/'NO CLICKS' of the detectors). Therefore, distinguishing between two ensembles requires the agent to point out a situation in which these ensembles give different predictions. By the rule of equation (35) we may calculate the probabilistic distribution of outcomes $i = 0, 1$ in a measurement for any distribution in the set conv $\mathcal{E}$. For a general mixture of equation (43), we get

$$
\begin{aligned}
P_{\vec{N}}(i) &= \sum_k p_k \, P_{\vec{N}_k}(i) \overset{(35)}{=} \sum_k p_k \, \tfrac{1}{2}\Big(1 + (-)^i \, \hat{z} \cdot \vec{N}_k\Big) \\
&\overset{(44)}{=} \tfrac{1}{2}\Big(1 + (-)^i \, \hat{z} \cdot \vec{N}\Big).
\end{aligned}
\tag{50}
$$

---

[7] These equivalence classes can be written in a more explicit form. For $|\vec{N}| = 1$, we have $[[\vec{N}]] =$ conv $[\vec{N}]$. If $|\vec{N}| < 1$, then $[[\vec{N}]]$ is a set-theoretic sum of sets $\sum_k p_k$ conv $[\vec{N}_k]$ over all possible convex decompositions of vector $\vec{N} = \sum_k p_k \, \vec{N}_k$ with $\vec{N}_k \in \boldsymbol{S}^2$.

This means that distributions in the same equivalence class $[[\vec{N}]]$ give identical probabilistic predictions which depend solely on the labelling vector $\vec{N} \in \boldsymbol{B}^3$ (actually only on its length $|\vec{N}|$ and polar angle $\theta$). We conclude that a measurement only by itself does not differentiate between distributions in the same class $[[\vec{N}]]$.

The only way for the agent to distinguish between two distributions in the same class would be to process them via a circuit which makes experimental predictions different. Let us check the behaviour of an ensemble described by a general distribution $\boldsymbol{p} \in [[\vec{N}]]$ as specified by equation (43). Clearly, we have $\vec{N}_{\boldsymbol{p}} = \vec{N}$. From the previous section we know that each gate in the circuit effects the corresponding transformation (see equations (41) and (42))

$$
\boldsymbol{p} \longrightarrow \boldsymbol{p}' = \sum_k p_k \, \boldsymbol{p}'_k,
\tag{51}
$$

where $\boldsymbol{p}'_k \in [\vec{N}_k']$ are specified by lemmas 1 and 2. For the purpose in hand it suffices to focus on the coarser level of classes which follow the rules of equations (34), (36) and (39). We get the following description of the associated vector

$$
\vec{N}_{\boldsymbol{p}'} = \sum_k p_k \, \vec{N}_k',
\tag{52}
$$

which is obtained after processing through the respective gates. For phase shifters $\mathbb{P}_j(\omega)$ and beam splitters $\mathbb{B}(\xi)$ it is given by

$$
\begin{aligned}
\vec{N}_{\boldsymbol{p}'} &\overset{(36)}{=} \sum_k p_k \, R_{\hat{r}}(\gamma)\vec{N}_k = R_{\hat{r}}(\gamma) \sum_k p_k \, \vec{N}_k \\
&\overset{(44)}{=} R_{\hat{r}}(\gamma)\vec{N}_{\boldsymbol{p}},
\end{aligned}
\tag{53}
$$

where $R_{\hat{r}}(\gamma)$ is the rotation about the axis $\hat{r} = \hat{z}$ or $\hat{x}$ corresponding to the respective gate (note that we have used linearity of rotations in $\mathbb{R}^3$). For detectors $\mathbb{D}_j$ by conditioning on the outcomes, we get

$$
\vec{N}_{\boldsymbol{p}'} \overset{(34)}{=} \sum_k p_k \, (\pm \hat{z}) = \pm \hat{z},
\tag{54}
$$

where the $\pm$ signs correspond to the respective outcomes (i.e. *real* particle in path $i = 0, 1$) which happen with the relative frequencies $P_{\vec{N}}(i)$ given in equation (50) (note that we have used the normalization condition $\sum_k p_k = 1$). Without conditioning on the outcomes, we have

$$
\begin{aligned}
\vec{N}_{\boldsymbol{p}'} &\overset{(39)}{=} \sum_k p_k \Big(P_{\vec{N}_k}(0)(+\hat{z}) + P_{\vec{N}_k}(1)(-\hat{z})\Big) \\
&\overset{(35)}{=} \sum_k p_k \Big(\tfrac{1}{2}(1 + \hat{z} \cdot \vec{N}_k) - \tfrac{1}{2}(1 - \hat{z} \cdot \vec{N}_k)\Big) \hat{z} \\
&\overset{(44)}{=} (\hat{z} \cdot \vec{N}) \, \hat{z}.
\end{aligned}
\tag{55}
$$

We conclude that whatever the action taken by the agent distributions from the same class $\boldsymbol{p}, \boldsymbol{q} \in [[\vec{N}]]$ transform into distributions again in the same class $\boldsymbol{p}', \boldsymbol{q}' \in [[\vec{N}']]$, i.e. we get $\vec{N}'_{\boldsymbol{p}} = \vec{N}'_{\boldsymbol{q}} = \vec{N}'$. As has already been noted this entails identical probabilistic predictions, see equation (50). In consequence, states in the same class $[[\vec{N}]]$ are operationally indistinguishable to the agent, which means that there is no way to make experimental predictions different by the means available to the agent. On the other hand, it is relatively easy to come up with a circuit which discriminates between distributions in different classes.[8]

We may briefly summarize the foregoing discussion in the following theorem.

**Theorem 1 (Ontic description of epistemic constraints).** *An agent subject to epistemic constraints explores the model defined in section 4 only to a limited extent. Whatever arrangement of gates in the circuit, possible conditioning on the outcomes and probabilistic mixing of ensembles, the agent remains confined within a restricted set of distributions. Her processing of the system via any conceivable circuit boils down to the following rules.*

---

[8] For distributions in different classes $[[\vec{N}_1]]$ and $[[\vec{N}_2]]$ it suffices, for example, to construct a circuit which 'rotates' one of the vectors to point north (if the vectors are of different length, take the longer one); such a circuit leads to a different distribution of outcomes as given by equation (50).

(i) *The range of distributions accessible to the agent is restricted to the set* $\mathrm{conv}\,\mathcal{E} \subset \mathcal{P}(\Omega)$ *which splits into equivalence classes* $[[\vec{N}]] \subset \mathcal{E}$ *labelled by vectors in the unit ball* $\vec{N} \in \boldsymbol{B}^3$. *Elements in the same class* $[[\vec{N}]]$ *are operationally indistinguishable by the means available to the agent.*

(ii) *Distributions in each class* $[[\vec{N}]]$ *transform congruently under any action of the agent, i.e. we have*

$$[[\vec{N}]] \ni \boldsymbol{p} \overset{\mathbb{T}}{\longrightarrow} \boldsymbol{p}' \in [[\vec{N}_{\mathbb{T}}]], \tag{56}$$

*where* $\vec{N}_{\mathbb{T}}$ *depends only on* $\vec{N}$ *and the kind of action* $\mathbb{T}$ *that is being performed. In such a case, mapping of classes is conveniently represented by the mapping of labelling vectors as explained below:*

*Phase shifters* $\mathbb{P}_j(\omega)$ *are described by rotations about the* $\hat{z}$-*axis*

$$\vec{N} \overset{\mathbb{P}_0(\omega)}{\longrightarrow} R_z(-\omega)\vec{N}, \tag{57}$$

$$\vec{N} \overset{\mathbb{P}_1(\omega)}{\longrightarrow} R_z(+\omega)\vec{N}. \tag{58}$$

*Beam splitters* $\mathbb{B}(\xi)$ *correspond to rotations about the* $\hat{x}$-*axis*

$$\vec{N} \overset{\mathbb{B}(\xi)}{\longrightarrow} R_x(\xi)\vec{N}. \tag{59}$$

*Detectors* $\mathbb{D}_j$ *implement measurements which consist in registering 'CLICKS' which reveal the position of the real particle in the circuit ('NO CLICK' in the case of a single detector* $\mathbb{D}_j$ *is a 'negative' measurement result which indicates the presence of the real particle in the other* $\bar{j}$ *th path ). For a system described by a distribution in class* $[[\vec{N}]]$ *the probability* $P_{\vec{N}}(i)$ *of outcome* $i = 0, 1$, *i.e. real particle is located in the ith path, is given by*

$$P_{\vec{N}}(i) = \tfrac{1}{2}\left(1 + (-)^i \,\hat{z} \cdot \vec{N}\right). \tag{60}$$

*Upon conditioning on the outcomes, post-measurement states in the selected subensembles are described by distributions in one of the respective classes* $[[\pm\hat{z}]]$, *i.e.*

$$\vec{N} \overset{\mathbb{D}_j}{\longrightarrow} \begin{cases} +\hat{z}, & \text{for outcome } i = 0, \\ -\hat{z}, & \text{for outcome } i = 1. \end{cases} \tag{61}$$

*If outcomes are not registered, then the whole ensemble after the measurement is described by*

$$\vec{N} \overset{\mathbb{D}_j}{\longrightarrow} (\hat{z} \cdot \vec{N})\,\hat{z}. \tag{62}$$

Observe that from the operational point of view a large part of description in theorem 1 is redundant. All information that is needed to make predictions about the behaviour of the system processed by any conceivable circuit is fully specified by the vector $\vec{N} \in \boldsymbol{B}^3$ which labels the class $[[\vec{N}]]$ in which the distribution is contained. Note that the characteristics of the classes in terms of vector $\vec{N}$ is *self-contained*, i.e. its use does not require further details concerning the underlying distribution. It is also *minimal* in a sense that different $\vec{N}$'s can be distinguished by appropriately chosen experiments. In this way, we get an operational framework which properly describes the model as seen by the agent without making reference to the underlying ontology.

**Corollary 1 (Operational account of the model).** *From the operational point of view the system is fully specified by a vector in the unit ball* $\vec{N} \in \boldsymbol{B}^3$. *Transformations effected by phase shifters* $\mathbb{P}_j(\omega)$ *and beam splitters* $\mathbb{B}(\xi)$ *correspond to rotations as given in equations* (57)–(59). *'CLICKS' of detectors* $\mathbb{D}_j$ *(measurements) are described by the probabilistic rule of equation* (60) *and conditioned on the outcomes of post-measurement states following the prescription of equation* (61) *(for non-selective measurement we have equation* (62)*).*

Note that in this formulation one is only concerned with 'CLICKS' of detectors in experiments and how the distributions of outcomes changes when processed via possible circuits. It does not even require the concept of a particle to interpret the results.[9] As a matter of fact, this is the only information which is relevant for the agent who is unaware or indifferent to the underlying ontology.

From the mathematical point of view the operational account of the model given in corollary 1 should be compared with description of a qubit in section 3. It is straightforward to see that the vector $\vec{N}$ plays the role of the Bloch vector $\boldsymbol{n}$ which follows the same rules describing transformations and measurements. This means that the mathematical framework of corollary 1 is identical to a qubit.

---

[9] Of course, if the agent thinks in terms of particles, then she may form a hypothesis that the system was prepared with a 'particle' in a given path. Then, the agent's concept of a 'particle' coincides with the *real* particle in the underlying ontological description. However, this naive picture is incomplete in missing out on the crucial element of the model that are *ghost* particles whose presence leads to nontrivial effects which by other means are hard to explain (see interpretative issues with the principle of locality in the quantum case).

To conclude, recall that from the construction the two-path quantum interferometric circuits (see section 3) and circuits built in our stochastic model (see section 4) are structurally the same. In this section we have shown that the operational descriptions are equivalent as well, i.e. in either case boil down to a qubit. This brings the final conclusion that from the agent's perspective quantum-interferometric circuits and their stochastic counterparts are indistinguishable.

## 6. Discussion

We have considered the question whether behaviour of a single quantum particle in two-path interferometric setups is enough to declare non-locality. If it were the case, this would mean the impossibility of a hidden-variable account based on particle ontology and the paradigm of locality. In this paper we express reservations towards statements to that effect. We have constructed an explicit stochastic model which simulates the behaviour of such quantum interferometric circuits with all particles and gates interacting strictly locally. It has been shown that from the agent's perspective predictions of the model are indistinguishable from the quantum case and its operational account is equivalent to a qubit. Thus, allegedly 'non-local' effects in the model are explained to arise only on the epistemic level of description by the agent whose knowledge is incomplete due to the restricted means of investigating the system. This shows that dismissing local realism only on the basis of arguments concerning a qubit in the interferometric setup is premature.

It should be noted that a similar conclusion can be drawn from the analysis of the de Broglie–Bohm interpretation of quantum mechanics [18, 19]. In this framework single-particle interferometric phenomena can be modelled as a particle guided by a local quantum potential (directly related with the wave function). Locality derives from the fact that in the single-particle case the configuration space on which the quantum potential is defined coincides with the real 3D space. As a result, non-locality can be explained by classical-like effects of the potential. A characteristic feature of the de Broglie-Bohm interpretation is that particles follow weird trajectories which lead to very complicated spatial descriptions and 'surrealistic' effects—also in the case of simple interferometric setups which are described as finite-dimensional systems in the relevant degrees of freedom (like those considered in this paper). We point out that the explanation suggested by the de Broglie–Bohm interpretation is different from the one offered by the model presented in this paper where particles follow the usual classical paths and there is no need for the potential in the model. However, there is an important similarity between both models which is the existence of some object in the other branch of the interferometer which encodes the information about the whole setup—this seems indispensable in any model explaining interferometric setups [20]. In the case of the de Broglie–Bohm model it is implemented by the presence of the potential (or empty wave) which fills the whole space, while in the case of our model it is the *ghost* particle traversing the other path which makes the information available only at the crossing points.

The simplicity of the interferometric realization of a qubit makes it an attractive framework for discussion of various paradoxes and quantum effects. For example, it is often taken as a prototypical situation illustrating quantum interference [4, 5], interaction-free measurements [6–8], the quantum Zeno effect [7–10], delayed-choice experiments [11, 12] or discussions of Leggett–Garg inequalities [13, 14]. In view of the presented model these kinds of arguments lose their original allure. All these effects have classical analogues which can be explained by incomplete knowledge and state disturbance, and hence are not reserved exclusively for the quantum realm. The main point of interest here is that single-particle effects in two-path quantum interferometric circuits can be simulated in a classical manner without resorting to non-locality.

Indeed, recent research show that many phenomena typically associated with strictly quantum mechanical effects have analogues in classical models with epistemic restrictions [21–29]. Most notable in this respect is the Spekkens' toy model [24] which reproduces a surprisingly large array of quantum phenomena in a simple discrete system constrained by the so-called 'knowledge balance principle' (it also reproduces certain aspects of the interferometric phenomena [30]). This idea has been taken further to a continuous model which reconstructs Gaussian quantum mechanics from the so-called Liouville mechanics of the classical phase-space subject to an epistemic restriction [25]. The main point of these models is that they are $\psi$-epistemic in the classification of reference [31]. Following these lines considerable effort has been taken towards understanding to what extent quantum states can be seen as states of knowledge and the possibility of reconstructing quantum theory based on this premise. However, there are strong results which suggest that it is not possible within the framework of $\psi$-epistemic theories [32, 33]. We note that our construction falls into the category of $\psi$-ontic models wherein these objections do not apply. Another important characteristic of ontological models which aim at reconstructing quantum predictions is their preparation, transformation and measurement contextuality [34–36]. The model presented in this paper has all these properties.[10]

---

[10] It is interesting to note that the model does not have more contextuality than is strictly required by the standard proofs. This is another difference with the de Broglie–Bohm interpretation in which results of a measurement in addition to the context of co-measured observables depend on the way in which a particular measurement is performed; see [37] for a discussion.

Let us remark that various ontological models of a qubit exist, e.g. models of Beltrametti–Bugajski [38], Bell–Mermin [1, 3] or Kochen–Specker [34]; see [31, 36] for a review. Their focus is, however, on the mathematical formalism rather than on the conceptual issues concerning the interpretation. In particular, it is not clear how to cast them into the framework of particles and interferometric circuits without violation of the locality principle. The problem lies in interpreting ontic states of these models as characteristics of local objects in order to avoid paradoxes associated with the collapse of the wave function. In this paper we take a different approach by settling the conceptual questions first. Here, we work from the outset with well-defined local objects (particles and gates) and show that the operational description of the model is equivalent to a qubit. In this way we get the correct mathematical formalism of a qubit and an unproblematic interpretation in terms of particles and gates which conform to the paradigm of locality. We point out that our model of a qubit fundamentally differs from the existing proposals which can be immediately observed by comparing the respective ontic state spaces.

A general conclusion from the paper is that single-particle effects in two-path interferometric circuits are not enough to establish non-locality. In other words, quantum non-locality is genuinely a multi-particle phenomenon, i.e. it requires at least two quantum particles to manifest nontrivial effects as considered in the Bell-type arguments. It would be interesting from the foundational point of view to complete the picture by considering the possibility of local simulation of a single quantum particle in a general multi-path interferometric setup which corresponds to a qudit [17]. We believe that this should be possible within the framework of ontological models with epistemic constraints.

## Appendix A. Proofs of lemmas 1 and 2

For the proof of lemmas 1 and 2 we should check behaviour of all distributions in $\mathcal{E} \subset \mathcal{P}(\Omega)$ under the action of the respective gates $\mathbb{P}_j(\omega)$, $\mathbb{B}(\xi)$ and $\mathbb{D}_j$. In the following, we first consider distributions in each class $[\vec{N}]$ defined in equations (18) and (19), i.e.

$$\boldsymbol{p}_{\vec{N}}^{(\alpha,\beta)} = \cos^2\left(\frac{\theta}{2}\right)\delta_0\ \delta_{R_z(\alpha)\vec{N}}\ \delta_\alpha\ +\ \sin^2\left(\frac{\theta}{2}\right)\delta_1\ \delta_{-R_z(\beta)\vec{N}}\ \delta_\beta\ \in[\vec{N}], \tag{63}$$

where $\vec{N} = (\theta, \phi) \in S^2$ and $\alpha,\ \beta \in [0, 2\pi)$. Then for completeness we have to examine the special case $[\pm\hat{z}]$ defined in equation (20). Here, it suffices to check only distributions in $[\pm z]^* \subset [\pm\hat{z}]$ defined in equation (21), i.e.

$$\delta_0\ \delta_{\vec{n}}\ \delta_\varnothing\in[+\hat{z}]^*\quad\text{and}\quad\delta_1\ \delta_{\vec{n}}\ \delta_\varnothing\in[-\hat{z}]^*, \tag{64}$$

where $\vec{n} = (\theta, \phi) \in S^2 \setminus \{-\hat{z}\}$. Note that the remaining distributions $\delta_0\ \delta_{\hat{z}}\ \delta_\alpha\in[+\hat{z}] \setminus [+z]^*$ and $\delta_1\ \delta_{\hat{z}}\ \delta_\beta\in[-\hat{z}] \setminus [-z]^*$ are precisely of the form considered in equation (63) for $\vec{N} = \pm\hat{z}$, and hence there is no need to check them again.

In the proofs we give the *explicit* action of the gates on each distribution $\boldsymbol{p}\in\mathcal{E}$. It is interesting to see how this unfolds in the case of detection; in particular, it explains the mechanism behind the purported non-local effects for 'negative' measurement results within the local framework of the model.

**Proof of lemma 1.** Let us first consider phase shifters $\mathbb{P}_j(\omega)$. According to definition 4 (i), their action is given by

$$[\vec{N}]\ni\boldsymbol{p}_{\vec{N}}^{(\alpha,\beta)}\ \xrightarrow{\mathbb{P}_0(\omega)}\ \boldsymbol{p}' =\ \cos^2\left(\frac{\theta}{2}\right)\delta_0\ \delta_{R_z(\alpha-\omega)\vec{N}}\ \delta_\alpha\ +\ \sin^2\left(\frac{\theta}{2}\right)\delta_1\ \delta_{-R_z(\beta)\vec{N}}\ \delta_{\beta+\omega}$$
$$=\ \boldsymbol{p}_{R_z(-\omega)\vec{N}}^{(\alpha,\beta')}\in[R_z(-\omega)\vec{N}], \tag{65}$$

where we have changed the variable $\beta' = \beta + \omega$, and

$$[\vec{N}]\ni\boldsymbol{p}_{\vec{N}}^{(\alpha,\beta)}\ \xrightarrow{\mathbb{P}_1(\omega)}\ \boldsymbol{p}' = \cos^2\left(\frac{\theta}{2}\right)\delta_0\ \delta_{R_z(\alpha)\vec{N}}\ \delta_{\alpha-\omega}\ +\ \sin^2\left(\frac{\theta}{2}\right)\delta_1\ \delta_{-R_z(\beta+\omega)\vec{N}}\ \delta_\beta$$
$$=\ \boldsymbol{p}_{R_z(\omega)\vec{N}}^{(\alpha',\beta)}\in[R_z(\omega)\vec{N}], \tag{66}$$

where we have changed the variable $\alpha' = \alpha - \omega$.

For distributions in $[\pm\hat{z}]^*$, using definition 4 (i), we get

$$[+\hat{z}]^*\ni\delta_0\ \delta_{\vec{n}}\ \delta_\varnothing\ \xrightarrow{\mathbb{P}_0(\omega)}\ \delta_0\ \delta_{R_z(-\omega)\vec{n}}\ \delta_\varnothing\in[+\hat{z}]^*, \tag{67}$$

$$[-\hat{z}]^*\ni\delta_1\ \delta_{\vec{n}}\ \delta_\varnothing\ \xrightarrow{\mathbb{P}_0(\omega)}\ \delta_1\ \delta_{\vec{n}}\ \delta_\varnothing\in[-\hat{z}]^*, \tag{68}$$

$$[+\hat{z}]^*\ni\delta_0\ \delta_{\vec{n}}\ \delta_\varnothing\ \xrightarrow{\mathbb{P}_1(\omega)}\ \delta_0\ \delta_{\vec{n}}\ \delta_\varnothing\in[+\hat{z}]^*, \tag{69}$$

$$[-\hat{z}]^*\ni\delta_1\ \delta_{\vec{n}}\ \delta_\varnothing\ \xrightarrow{\mathbb{P}_1(\omega)}\ \delta_1\ \delta_{R_z(\omega)\vec{n}}\ \delta_\varnothing\in[-\hat{z}]^*. \tag{70}$$

Now, we consider the case of beam splitters $\mathbb{B}(\xi)$. From definition 4 (ii), we have

$$
\begin{aligned}
[\vec{N}] \ni \boldsymbol{p}_{\vec{N}}^{(\alpha, \beta)} \xrightarrow{\mathbb{B}(\xi)} \boldsymbol{p}' &= \cos^2\left(\frac{\theta}{2}\right)\left\{\cos^2\left(\frac{\theta'}{2}\right)\delta_0\,\delta_{\vec{N}'}\,\delta_0 \;+\; \sin^2\left(\frac{\theta'}{2}\right)\delta_1\,\delta_{-\vec{N}'}\,\delta_0\right\} \\
&\quad + \sin^2\left(\frac{\theta}{2}\right)\left\{\cos^2\left(\frac{\tilde{\theta}'}{2}\right)\delta_1\,\delta_{\tilde{N}'}\,\delta_0 \;+\; \sin^2\left(\frac{\tilde{\theta}'}{2}\right)\delta_0\,\delta_{-\tilde{N}'}\,\delta_0\right\} \\
&\overset{(*)}{=} \cos^2\left(\frac{\theta}{2}\right)\left\{\cos^2\left(\frac{\theta'}{2}\right)\delta_0\,\delta_{\vec{N}'}\,\delta_0 \;+\; \sin^2\left(\frac{\theta'}{2}\right)\delta_1\,\delta_{-\vec{N}'}\,\delta_0\right\} \\
&\quad + \sin^2\left(\frac{\theta}{2}\right)\left\{\sin^2\left(\frac{\theta'}{2}\right)\delta_1\,\delta_{-\vec{N}'}\,\delta_0 \;+\; \cos^2\left(\frac{\theta'}{2}\right)\delta_0\,\delta_{\vec{N}'}\,\delta_0\right\} \\
&= \cos^2\left(\frac{\theta'}{2}\right)\delta_0\,\delta_{\vec{N}'}\,\delta_0 \;+\; \sin^2\left(\frac{\theta'}{2}\right)\delta_1\,\delta_{-\vec{N}'}\,\delta_0 \\
&= \boldsymbol{p}_{R_x(\xi)\vec{N}}^{(0,0)} \in [R_x(\xi)\vec{N}],
\end{aligned}
\tag{71}
$$

where

$$
\begin{aligned}
\vec{N}' &= (\theta', \phi') = R_x(\xi)\,R_z(-\alpha)\,R_z(\alpha)\vec{N} = R_x(\xi)\vec{N}, \\
\tilde{\vec{N}}' &= (\tilde{\theta}', \tilde{\phi}') = R_x(\xi)\,R_z(-\beta)\left(-R_z(\beta)\vec{N}\right) = -R_x(\xi)\vec{N}.
\end{aligned}
$$

The second equality (*) is a straightforward consequence of $\tilde{\vec{N}}' = -\vec{N}'$. In particular, this entails $\tilde{\theta}' = \pi - \theta'$, which allows for substitution: $\cos^2\left(\frac{\tilde{\theta}'}{2}\right) = \sin^2\left(\frac{\theta'}{2}\right)$ and $\sin^2\left(\frac{\tilde{\theta}'}{2}\right) = \cos^2\left(\frac{\theta'}{2}\right)$. The rest boils down to collecting the terms which simplify by the use of elementary trigonometric identity: $\cos^2\left(\frac{\theta}{2}\right) + \sin^2\left(\frac{\theta}{2}\right) = 1$.

Finally, for distributions in $[\pm\hat{z}]^*$, using definition 4 (ii), we get

$$
[+\hat{z}]^* \ni \delta_0\,\delta_{\vec{n}}\,\delta_\varnothing \xrightarrow{\mathbb{B}(\xi)} \boldsymbol{p}' = \cos^2\left(\frac{\theta'}{2}\right)\delta_0\,\delta_{\vec{n}'}\,\delta_0 \;+\; \sin^2\left(\frac{\theta'}{2}\right)\delta_1\,\delta_{-\vec{n}'}\,\delta_0
$$
$$
= \boldsymbol{p}_{R_x(\xi)\hat{z}}^{(0,0)} \in [R_x(\xi)\hat{z}],
\tag{72}
$$

where $\vec{n}' = (\theta', \phi') = R_x(\xi)\hat{z}$, and

$$
\begin{aligned}
[-\hat{z}]^* \ni \delta_1\,\delta_{\vec{n}}\,\delta_\varnothing \xrightarrow{\mathbb{B}(\xi)} \boldsymbol{p}' &= \cos^2\left(\frac{\theta'}{2}\right)\delta_1\,\delta_{\vec{n}'}\,\delta_0 \;+\; \sin^2\left(\frac{\theta'}{2}\right)\delta_0\,\delta_{-\vec{n}'}\,\delta_0 \\
&\overset{(**)}{=} \sin^2\left(\frac{\tilde{\theta}'}{2}\right)\delta_1\,\delta_{-\vec{n}'}\,\delta_0 \;+\; \cos^2\left(\frac{\tilde{\theta}'}{2}\right)\delta_0\,\delta_{\tilde{n}'}\,\delta_0 \\
&= \boldsymbol{p}_{-R_x(\xi)\hat{z}}^{(0,0)} \in [-R_x(\xi)\hat{z}],
\end{aligned}
\tag{73}
$$

where $\vec{n}' = (\theta', \phi') = R_x(\xi)\hat{z}$. In the second equality (**) we have changed the variable $\tilde{\vec{n}}' = -\vec{n}'$ and consequently substituted: $\cos^2\left(\frac{\theta'}{2}\right) = \sin^2\left(\frac{\tilde{\theta}'}{2}\right)$ and $\sin^2\left(\frac{\theta'}{2}\right) = \cos^2\left(\frac{\tilde{\theta}'}{2}\right)$; see the same trick used in justifying equality (*) above.

In this way, we have checked action of phase shifters $\mathbb{P}_j(\omega)$ and beam splitters $\mathbb{B}(\xi)$ on each distribution $\boldsymbol{p} \in \mathcal{E}$, which concludes the proof of lemma 1. $\qquad\square$

**Proof of lemma 2.** First we will consider the case of a single detector $\mathbb{D}_j$ placed in the $j$th path. The detector 'Clicks' only if there is a *real* particle in the $j$th path and 'No Click' testifies to the presence of the *real* particle in the other $\bar{j}$th path, see definition 4 (iii). Therefore, for all distributions $\boldsymbol{p} \in [\vec{N}]$ with $\vec{N} = (\theta, \phi) \in S^2$ the probability of the outcome corresponding to the *real* particle being in path $i = 0, 1$ is given by the respective coefficients in equations (63) and (64), i.e.

$$
\begin{aligned}
P_{\vec{N}}(i = 0) &= \cos^2\left(\frac{\theta}{2}\right) = \tfrac{1}{2}(1 + \cos\theta) = \tfrac{1}{2}(1 + \hat{z}\cdot\vec{N}), \\
P_{\vec{N}}(i = 1) &= \sin^2\left(\frac{\theta}{2}\right) = \tfrac{1}{2}(1 - \cos\theta) = \tfrac{1}{2}(1 - \hat{z}\cdot\vec{N}),
\end{aligned}
\tag{74}
$$

where we have used the fact that $\cos\theta = \hat{z}\cdot\vec{N}$. In particular, we have $P_{+\hat{z}}(0) = P_{-\hat{z}}(1) = 1$ and $P_{+\hat{z}}(1) = P_{-\hat{z}}(0) = 0$.

We observe that according to definition 4 (iii) state after detection depends on the outcome. By conditioning on whether the detector 'Clicked' or did not 'Click' we get the following description of post-measurement states. In the case of a single detector $\mathbb{D}_0$, we get

$$[\vec{N}] \ni \boldsymbol{p}_{\vec{N}}^{(\alpha,\beta)} \xrightarrow{\mathbb{D}_0} \begin{cases} \delta_0 \; \delta_{\hat{z}} \; \delta_{\alpha} \;\; \in [+\hat{z}], \quad \text{if detector 'CLICKED',} \\ \hspace{3.5cm} \text{\scriptsize happens with } P_{\vec{N}}(i=0) \\ \delta_1 \; \delta_{-\vec{N}} \; \delta_{\varnothing} \in [-\hat{z}]^*, \quad \text{if detector did not 'CLICK',} \\ \hspace{3.5cm} \text{\scriptsize happens with } P_{\vec{N}}(i=1) \end{cases} \tag{75}$$

and for the remaining distributions in $[\pm\hat{z}]^*$, we have

$$[+\hat{z}]^* \ni \delta_0 \; \delta_{\vec{n}} \; \delta_{\varnothing} \xrightarrow{\mathbb{D}_0} \delta_0 \; \delta_{\hat{z}} \; \delta_{\varnothing} \in [+\hat{z}]^*, \qquad \text{detector always 'CLICKS',} \tag{76}$$

$$[-\hat{z}]^* \ni \delta_1 \; \delta_{\vec{n}} \; \delta_{\varnothing} \xrightarrow{\mathbb{D}_0} \delta_1 \; \delta_{\vec{n}} \; \delta_{\varnothing} \in [-\hat{z}]^*, \qquad \text{detector never 'CLICKS'.} \tag{77}$$

Similarly, for a single detector $\mathbb{D}_1$, we get

$$[\vec{N}] \ni \boldsymbol{p}_{\vec{N}}^{(\alpha,\beta)} \xrightarrow{\mathbb{D}_1} \begin{cases} \delta_0 \; \delta_{\vec{N}} \; \delta_{\varnothing} \;\; \in [+\hat{z}]^*, \quad \text{if detector did not 'CLICK',} \\ \hspace{3.5cm} \text{\scriptsize happens with } P_{\vec{N}}(i=0) \\ \delta_1 \; \delta_{\hat{z}} \; \delta_{\beta} \;\; \in [-\hat{z}], \quad \text{if detector 'CLICKED',} \\ \hspace{3.5cm} \text{\scriptsize happens with } P_{\vec{N}}(i=1) \end{cases} \tag{78}$$

and

$$[+\hat{z}]^* \ni \delta_0 \; \delta_{\vec{n}} \; \delta_{\varnothing} \xrightarrow{\mathbb{D}_1} \delta_0 \; \delta_{\vec{n}} \; \delta_{\varnothing} \in [+\hat{z}]^*, \qquad \text{detector never 'CLICKS',} \tag{79}$$

$$[-\hat{z}]^* \ni \delta_1 \; \delta_{\vec{n}} \; \delta_{\varnothing} \xrightarrow{\mathbb{D}_1} \delta_1 \; \delta_{\hat{z}} \; \delta_{\varnothing} \in [-\hat{z}]^*, \qquad \text{detector always 'CLICKS'.} \tag{80}$$

(We remark that for $\vec{N} = +\hat{z}$ in equation (75) and $\vec{N} = -\hat{z}$ in equation (78) measurements with 'negative' results do not lead to distributions $\delta_1 \; \delta_{-\hat{z}} \; \delta_{\varnothing}$ and $\delta_0 \; \delta_{-\hat{z}} \; \delta_{\varnothing}$ which have been meticuluously excluded in the definition of $[\pm\hat{z}]^*$ in equation (21), since they happen with zero probability.)

Finally, in the case when detectors are placed in both paths $\mathbb{D}_1$ and $\mathbb{D}_2$ their outcomes are perfectly *anti-correlated*, i.e. if one 'CLICKS' the other does not ('NO CLICK'). This is because there is a single *real* particle in the system (either in path $i = 0$ or path $i = 1$). Clearly, probabilities of outcomes follow the pattern of equation (74) and conditioned on the outcomes post-measurement states are given by

$$[\vec{N}] \ni \boldsymbol{p}_{\vec{N}}^{(\alpha,\beta)} \xrightarrow{\mathbb{D}_0 \; \& \; \mathbb{D}_1} \begin{cases} \delta_0 \; \delta_{\hat{z}} \; \delta_{\varnothing} \;\; \in [+\hat{z}]^*, \quad \text{if detector } \mathbb{D}_0 \text{ 'CLICKED',} \\ \hspace{3.5cm} \text{\scriptsize happens with } P_{\vec{N}}(i=0) \\ \delta_1 \; \delta_{\hat{z}} \; \delta_{\varnothing} \;\; \in [-\hat{z}]^*, \quad \text{if detector } \mathbb{D}_1 \text{ 'CLICKED',} \\ \hspace{3.5cm} \text{\scriptsize happens with } P_{\vec{N}}(i=1) \end{cases} \tag{81}$$

and

$$[+\hat{z}]^* \ni \delta_0 \; \delta_{\vec{n}} \; \delta_{\varnothing} \xrightarrow{\mathbb{D}_0 \; \& \; \mathbb{D}_1} \delta_0 \; \delta_{\hat{z}} \; \delta_{\varnothing} \in [+\hat{z}]^*, \qquad \text{only detector } \mathbb{D}_0 \text{ 'CLICKS',} \tag{82}$$

$$[-\hat{z}]^* \ni \delta_1 \; \delta_{\vec{n}} \; \delta_{\varnothing} \xrightarrow{\mathbb{D}_0 \; \& \; \mathbb{D}_1} \delta_1 \; \delta_{\hat{z}} \; \delta_{\varnothing} \in [-\hat{z}]^*, \qquad \text{only detector } \mathbb{D}_1 \text{ 'CLICKS'.} \tag{83}$$

This exhausts all distributions in the restricted set $\mathcal{E}$ and concludes the proof of lemma 2. $\qquad \square$

## Appendix B. Proof of proposition 1 (ii)

There are many ways to reach the given distribution $\boldsymbol{p} \in \mathcal{E}$ starting from the initial states $\boldsymbol{p}_i^{(\text{in})}$ in equation (17). For the proof of proposition 1 (ii) it suffices to give an example of such a protocol for each $\boldsymbol{p} \in \mathcal{E}$. In the following we rely on the geometrical picture of transformations (rotations and 'projections') and their explicit form given in the proofs of lemmas 1 and 2.

Let us start with a procedure which leads to distributions of the form $\delta_0 \; \delta_{\vec{n}} \; \delta_{\varnothing} \in [+\hat{z}]^*$ with $\vec{n} = (\theta, \phi) \in S^2 \setminus \{-\hat{z}\}$, see equation (21). We first apply to the initial distribution $\boldsymbol{p}_0^{(\text{in})} \equiv \delta_0 \; \delta_{\hat{z}} \; \delta_{\varnothing}$ a sequence of gates which implement the following two rotations $R_z(\phi + \frac{\pi}{2}) R_x(\theta)\hat{z} = \vec{n}$ using the beam splitter $\mathbb{B}(\theta)$ and phase shifter $\mathbb{P}_1(\phi + \frac{\pi}{2})$. Then we make a 'negative' measurement with a single detector $\mathbb{D}_1$ which consists in selecting the subensemble associated with 'NO CLICK' in the detector. This gives the following chain of transformations

$$\boldsymbol{p}_0^{(\text{in})} \xrightarrow[\text{(72)}]{\mathbb{B}(\theta)} \boldsymbol{p}_{R_x(\theta)\hat{z}}^{(0,0)} \xrightarrow[\text{(66)}]{\mathbb{P}_1(\phi + \frac{\pi}{2})} \boldsymbol{p}_{R_z(\phi + \frac{\pi}{2})R_x(\theta)\hat{z}}^{(-\phi - \frac{\pi}{2}, 0)} = \boldsymbol{p}_{\vec{n}}^{(-\phi - \frac{\pi}{2}, 0)}$$

$$\xrightarrow[\text{(78)}]{\mathbb{D}_1} \delta_0 \; \delta_{\vec{n}} \; \delta_{\varnothing}, \qquad \text{if detector } \mathbb{D}_1 \text{ did not 'CLICK'.} \tag{84}$$

Since $\vec{n} \neq -\hat{z}$, selection in the last step takes place with non-zero probability $P_{\vec{N}}(i = 0) = \frac{1}{2}(1 + \hat{z} \cdot \vec{n}) > 0$. This is actually the reason behind exclusion of this 'singular' point in the definition of $[+\hat{z}]^*$ in equation (21) (i.e. the agent has no way to prepare state $\delta_0 \; \delta_{-\hat{z}} \; \delta_{\varnothing} \in \mathcal{P}(\Omega)$; see also the remark after equation (80)).

In a similar manner one prepares distributions $\delta_1 \; \delta_{\vec{n}} \; \delta_{\oslash} \in [-\hat{z}]^*$ with $\vec{n} = (\theta, \phi) \in S^2 \setminus \{-\hat{z}\}$, see equation (21). Here, we first implement two rotations $R_z(\phi - \frac{\pi}{2})R_x(\pi - \theta)\hat{z} = -\vec{n}$ via gates $\mathbb{B}(\pi - \theta)$ and $\mathbb{P}_1(\phi - \frac{\pi}{2})$, and then condition on the 'negative' outcome in detector $\mathbb{D}_0$. We get

$$\boldsymbol{p}_0^{(\mathrm{in})} \xrightarrow[(72)]{\mathbb{B}(\pi-\theta)} \boldsymbol{p}_{R_x(\pi-\theta)\hat{z}}^{(0,0)} \xrightarrow[(66)]{\mathbb{P}_1(\phi-\frac{\pi}{2})} \boldsymbol{p}_{R_z(\phi-\frac{\pi}{2})R_x(\pi-\theta)\hat{z}}^{(-\phi+\frac{\pi}{2},0)} = \boldsymbol{p}_{-\vec{n}}^{(-\phi+\frac{\pi}{2},0)}$$

$$\xrightarrow[(75)]{\mathbb{D}_0} \delta_1 \; \delta_{\vec{n}} \; \delta_{\oslash}, \qquad \text{if detector } \mathbb{D}_0 \text{ did not 'CLICK'.} \tag{85}$$

Here, selection in the last step takes place with non-zero probability as well $P_{\vec{N}}(i=1) = \frac{1}{2}(1 - \hat{z} \cdot (-\vec{n})) > 0$. It is guaranteed by exclusion of the point $\vec{n} = -\hat{z}$ in the definition of $[-\hat{z}]^*$ in equation (21) (i.e. the agent can not reach state $\delta_1 \; \delta_{-\hat{z}} \; \delta_{\oslash} \in \mathcal{P}(\Omega)$; see also the remark after equation (80)).

Now, we give a procedure which leads to distributions of the form $\boldsymbol{p}_{\vec{N}}^{(\alpha,\beta)} \in [\vec{N}]$ with $\vec{N} = (\theta, \phi) \in S^2$, see equation (19). First we note that any vector $\vec{N}$ can be obtained from $\hat{z}$ by appropriate rotations. Furthermore, the latter can be chosen as a sequence of rotations about axes $\hat{x}$ and $\hat{z}$ in a particular form $R_x(\xi_2)R_z(\omega)R_x(\xi_1)\hat{z} = \vec{N}$ for some choice of algles $\xi_1$, $\xi_2$ and $\omega$. If we implement rotations by the corresponding gates $\mathbb{B}(\xi_1)$, $\mathbb{P}_1(\omega)$ and $\mathbb{B}(\xi_2)$, we get the following description

$$\boldsymbol{p}_0^{(\mathrm{in})} \xrightarrow[(72)]{\mathbb{B}(\xi_1)} \boldsymbol{p}_{R_x(\xi_1)\hat{z}}^{(0,0)} \xrightarrow[(66)]{\mathbb{P}_1(\omega)} \boldsymbol{p}_{R_z(\omega)R_x(\xi_1)\hat{z}}^{(-\omega,0)} \xrightarrow[(71)]{\mathbb{B}(\xi_2)} \boldsymbol{p}_{R_x(\xi_2)R_z(\omega)R_x(\xi_1)\hat{z}}^{(0,0)} = \boldsymbol{p}_{\vec{N}}^{(0,0)}. \tag{86}$$

In this way, the agent prepares the distribution $\boldsymbol{p}_{\vec{N}}^{(0,0)}$ for any $\vec{N} \in S^2$. In order to get $\boldsymbol{p}_{\vec{N}}^{(\alpha,\beta)}$ with arbitrary $\alpha, \beta \in [0, 2\pi)$ it suffices that the agent first prepares $\boldsymbol{p}_{\vec{N}'}^{(0,0)}$ with $\vec{N}' = R_z(\alpha + \beta)\vec{N}$ and then applies phase shifters $\mathbb{P}_0(\beta)$ and $\mathbb{P}_1(-\alpha)$ in the respective paths, which gives

$$\boldsymbol{p}_{\vec{N}'}^{(0,0)} \xrightarrow[(65)]{\mathbb{P}_0(\beta)} \boldsymbol{p}_{R_z(-\beta)\vec{N}'}^{(0,\beta)} \xrightarrow[(66)]{\mathbb{P}_1(-\alpha)} \boldsymbol{p}_{R_z(-\alpha-\beta)\vec{N}'}^{(\alpha,\beta)} = \boldsymbol{p}_{\vec{N}}^{(\alpha,\beta)}. \tag{87}$$

In conclusion, we have given examples of protocols by which the agent prepares ensembles described by any distribution $\boldsymbol{p} \in \mathcal{E}$ from the initial ensemble $\boldsymbol{p}_0^{(\mathrm{in})}$ using an appropriate sequence of gates (for $\boldsymbol{p}_1^{(\mathrm{in})}$ proofs are analogous).

## Appendix C. Proof of proposition 3

Suppose that we have two different decompositions of a mixture $\boldsymbol{p} \in \mathrm{conv}\,\mathcal{E}$, i.e. $\boldsymbol{p} = \sum_k p_k \; \boldsymbol{p}_k = \sum_l q_l \; \boldsymbol{q}_l$ with $\boldsymbol{p}_k \in [\vec{N}_k]$ and $\boldsymbol{q}_l \in [\vec{N}_l']$. In a more explicit form it is written as follows (see equations (18)–(21))

$$\boldsymbol{p} = \sum_{\vec{N},\alpha,\beta} p_{\vec{N}}^{(\alpha,\beta)} \; \boldsymbol{p}_{\vec{N}}^{(\alpha,\beta)} \; + \sum_{i,\gamma} p_{i,\gamma} \; \delta_i \; \delta_{\hat{z}} \; \delta_{\gamma} \; + \sum_{i,\vec{n}} p_{i,\vec{n}} \; \delta_i \; \delta_{\vec{n}} \; \delta_{\oslash}, \tag{88}$$

$$\boldsymbol{p} = \sum_{\vec{N},\alpha,\beta} q_{\vec{N}}^{(\alpha,\beta)} \; \boldsymbol{p}_{\vec{N}}^{(\alpha,\beta)} \; + \sum_{i,\gamma} q_{i,\gamma} \; \delta_i \; \delta_{\hat{z}} \; \delta_{\gamma} \; + \sum_{i,\vec{n}} q_{i,\vec{n}} \; \delta_i \; \delta_{\vec{n}} \; \delta_{\oslash}, \tag{89}$$

with indices ranging over $\vec{N} \in S^2 \setminus \{\pm \hat{z}\}$, $\alpha, \beta, \gamma \in [0, 2\pi)$, $i \in \{0, 1\}$, $\vec{n} \in S^2 \setminus \{-\hat{z}\}$, and a finite number of non-zero coefficients $p_{\vec{N}}^{(\alpha,\beta)}$, $p_{i,\gamma}$, $p_{i,\vec{n}} \geqslant 0$ and $q_{\vec{N}}^{(\alpha,\beta)}$, $q_{i,\gamma}$, $q_{i,\vec{n}} \geqslant 0$ satisfying the usual normalization conditions

$$\sum_{\vec{N},\alpha,\beta} p_{\vec{N}}^{(\alpha,\beta)} + \sum_{i,\gamma} p_{i,\gamma} + \sum_{i,\vec{n}} p_{i,\vec{n}} = 1, \tag{90}$$

$$\sum_{\vec{N},\alpha,\beta} q_{\vec{N}}^{(\alpha,\beta)} + \sum_{i,\gamma} q_{i,\gamma} + \sum_{i,\vec{n}} q_{i,\vec{n}} = 1. \tag{91}$$

Since we have two different decompositions, then following equation (44) we can associate two vectors to $\boldsymbol{p}$, i.e.

$$\vec{N}_{\boldsymbol{p}} = \sum_{\vec{N}} \sum_{\alpha,\beta} p_{\vec{N}}^{(\alpha,\beta)} \; \vec{N} \; + \sum_{i,\gamma} (-)^i \, p_{i,\gamma} \; \hat{z} \; + \sum_{i,\vec{n}} (-)^i \, p_{i,\vec{n}} \; \hat{z}, \tag{92}$$

$$\vec{N}_{\boldsymbol{p}}' = \sum_{\vec{N}} \sum_{\alpha,\beta} q_{\vec{N}}^{(\alpha,\beta)} \; \vec{N} \; + \sum_{i,\gamma} (-)^i \, q_{i,\gamma} \; \hat{z} \; + \sum_{i,\vec{n}} (-)^i \, q_{i,\vec{n}} \; \hat{z}. \tag{93}$$

We need to prove uniqueness, which means that these vectors are the same $\vec{N}_{\boldsymbol{p}} = \vec{N}_{\boldsymbol{p}}'$.

Let us start by writing out the supports of the distributions which define these decompositions (see equations (19)–(21))

$$\text{supp } \delta_i \, \delta_{\vec{n}} \, \delta_\varnothing = \big\{ (i, \, \vec{n}, \, \varnothing) \big\},$$
$$\text{supp } \delta_i \, \delta_{\hat{z}} \, \delta_\gamma = \big\{ (i, \, \hat{z}, \, \gamma) \big\},$$
$$\text{supp } \boldsymbol{p}_{\vec{N}}^{(\alpha, \beta)} = \big\{ (0, \, R_z(\alpha) \vec{N}, \, \alpha), \, (1, \, -R_z(\beta) \vec{N}, \, \beta) \big\}. \tag{94}$$

Clearly, equations (88) and (89) are decompositions of the same function $\boldsymbol{p} : \Omega \to [0, 1]$. This means that we can compare values at each point $\omega \in \Omega$ obtained in both ways. By virtue of equation (94), we get

$$\boldsymbol{p}((i, \, \vec{n}, \, \varnothing)) \overset{(88)}{=} p_{i, \vec{n}} \overset{(89)}{=} q_{i, \vec{n}} \qquad \text{for all } i = 0, 1 \text{ and } \vec{n} \in S^2 \setminus \{-\hat{z}\}, \tag{95}$$

$$\boldsymbol{p}((i, \, \hat{z}, \, \gamma)) \overset{(88)}{=} p_{i, \gamma} \overset{(89)}{=} q_{i, \gamma} \qquad \text{for all } i = 0, 1 \text{ and } \gamma \in [0, 2\pi), \tag{96}$$

$$\boldsymbol{p}((0, \, \vec{M}, \, \alpha)) \overset{(88)}{=} \sum_{\substack{\vec{N}', \alpha', \beta' \text{ such that} \\ \vec{M} = R_z(\alpha') \vec{N}' \text{ and } \alpha' = \alpha}} p_{\vec{N}'}^{(\alpha', \beta')} = \sum_{\beta'} p_{R_z(-\alpha) \vec{M}}^{(\alpha, \beta')}$$

$$\overset{(89)}{=} \sum_{\substack{\vec{N}', \alpha', \beta' \text{ such that} \\ \vec{M} = R_z(\alpha') \vec{N}' \text{ and } \alpha' = \alpha}} q_{\vec{N}'}^{(\alpha', \beta')} = \sum_{\beta'} q_{R_z(-\alpha) \vec{M}}^{(\alpha, \beta')}$$

$$\text{for all } \vec{M} \in S^2 \setminus \{\pm \hat{z}\} \text{ and } \alpha \in [0, 2\pi). \tag{97}$$

Since $\vec{M}$ and $\alpha$ are arbitrary, by taking $\vec{M} = R_z(\alpha) \vec{N}$ and comparing the right-hand sides of equation (97), we get

$$\sum_{\beta} p_{\vec{N}}^{(\alpha, \beta)} = \sum_{\beta} q_{\vec{N}}^{(\alpha, \beta)} \qquad \text{for all } \vec{N} \in S^2 \setminus \{\pm \hat{z}\} \text{ and } \alpha \in [0, 2\pi). \tag{98}$$

Summation over $\alpha$ gives

$$\sum_{\alpha, \beta} p_{\vec{N}}^{(\alpha, \beta)} = \sum_{\alpha, \beta} q_{\vec{N}}^{(\alpha, \beta)} \qquad \text{for all } \vec{N} \in S^2 \setminus \{\pm \hat{z}\}. \tag{99}$$

Comparison of equations (92) and (93) with help of equalities equations (95), (96) and (99) shows that both vectors are the same $\vec{N}_{\boldsymbol{p}} = \vec{N}_{\boldsymbol{p}}'$, which concludes the proof.